\documentclass[twoside,leqno,twocolumn]{article}

\usepackage[letterpaper]{geometry}

\usepackage{ltexpprt}
\usepackage{xspace}
\usepackage{thmtools, thm-restate} 
\usepackage{amsmath}

\newcommand\blfootnote[1]{%
    \begingroup
    \renewcommand\thefootnote{}\footnote{#1}%
    \addtocounter{footnote}{-1}%
    \endgroup
}

\input{commands}

\begin{document}

\title{A Constant Factor Approximation for Navigating Through Connected Obstacles in the Plane\thanks{Kumar and Suri are 
supported by NSF award CCF-1814172. Lokshtanov is supported by BSF grant no. 2018302 and NSF award CCF-2008838.}}

\author{
Neeraj Kumar\thanks{Department of Computer Science, University of California, Santa Barbara, USA {\sf neeraj@cs.ucsb.edu, daniello@ucsb.edu, suri@cs.ucsb.edu}} 
\and
Daniel Lokshtanov\footnotemark[2]
\and
Saket Saurabh\thanks{Institute of Mathematical Sciences, Chennai, India and Department of Informatics, University of Bergen, Norway. {\sf saket@imsc.res.in} } 
\and
Subhash Suri\footnotemark[2]
}

\date{}

\maketitle

\fancyfoot[R]{\scriptsize{Copyright \textcopyright\ 2021 by SIAM\\
Unauthorized reproduction of this article is prohibited}}
\blfootnote{
    \hspace{-2.5em}
    \newline
    \begin{minipage}{0.33\textwidth}
        \footnotesize
        Saket Saurabh is supported by funding from the European Research Council
        (ERC) under the European Union's Horizon 2020 research and innovation
        programme (grant agreement No. 819416) and also  acknowledges the support of Swarnajayanti Fellowship grant  DST/SJF/MSA-01/2017-18. 
    \end{minipage}
    \hspace{0.5em}
    \begin{minipage}{0.1\textwidth}
        \begin{center}
            \includegraphics[scale=0.15]{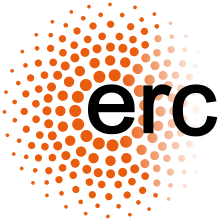}
            \hspace{0.2em}
            \includegraphics[scale=0.05]{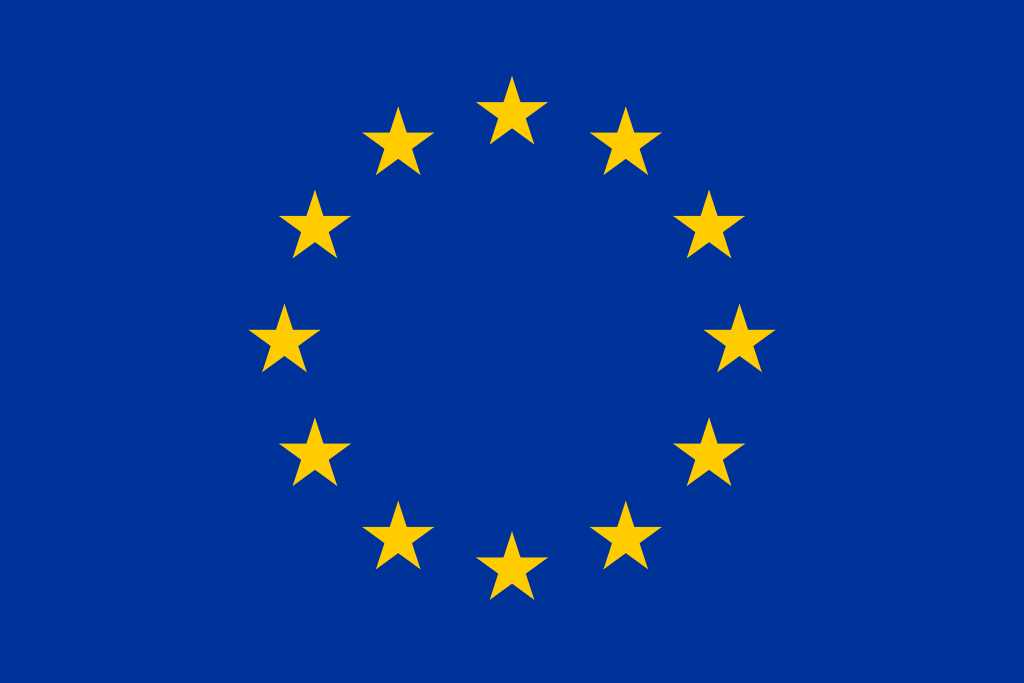}
        \end{center}
    \end{minipage}
}



\begin{abstract} \small\baselineskip=9pt
Given two points $s$ and $t$ in the plane and a set of obstacles defined by closed curves, what is the minimum number of obstacles 
touched by a path connecting $s$ and $t$? This is a fundamental and well-studied problem arising naturally in computational geometry, 
graph theory (under the names \mcp and {\sc Minimum Label Path}), wireless sensor networks ({\sc Barrier Resilience}) and motion 
planning ({\sc Minimum Constraint Removal}). It remains \textsf{NP}-hard even for very simple-shaped obstacles such as unit-length 
line segments. In this paper we give the first constant factor approximation algorithm for this problem, resolving an open problem 
of [Chan and Kirkpatrick, TCS, 2014] and [Bandyapadhyay et al., CGTA, 2020]. We also 
obtain a constant factor approximation for the {\sc Minimum Color Prize Collecting Steiner Forest} where the goal is to connect 
multiple request pairs $(s_1, t_1), \ldots, (s_k, t_k)$ while minimizing the number of obstacles touched by any $(s_i,t_i)$ path 
plus a fixed cost of $w_i$ for each pair $(s_i,t_i)$ left disconnected.  This generalizes the classic {\sc Steiner Forest} and 
{\sc Prize-Collecting Steiner Forest} problems on planar graphs, for which intricate PTASes are known. In contrast, no PTAS is 
possible for \mcp even on planar graphs since the problem is known to be APX-hard~[Eiben and Kanj, TALG, 2020].
Additionally, we show that generalizations of the problem to disconnected obstacles in the plane or connected obstacles in higher 
dimensions are strongly inapproximable assuming some well-known hardness conjectures.
\end{abstract}

\section{Introduction} \label{sec:intro}
We consider the following generalization of the undirected, unweighted $s$-$t$-{\sc Shortest Path}  problem, called  \mcp. 
Given an undirected graph $G = (V,E)$ with a coloring function $\sigma: V \rightarrow 2^{[m]}$ associated with its vertices, where 
$[m] = \{1, 2, \dots, m\}$ denotes the set of colors and $2^{[m]}$ denotes the family of all possible subsets of $[m]$, the
goal is to find an $s$-$t$ path $\pi$ that minimizes the number of distinct colors it touches. Formally, given a path $\pi$ in $G$, 
if $\sigma(\pi) = \bigcup_{v \in \pi} \sigma(v)$ is set of colors on the path, then the goal is to find an $s$-$t$ path $\pi$ 
minimizing $|\sigma(\pi)|$.
In other words, if each color represents an obstacle, then the \mcp problem for a $s$-$t$ path with the least number of obstacles on it.
This is a well-studied problem, first introduced by Jacob et al.~\cite{jacob1999minimum}, and independently re-discovered and studied 
in several application domains, including wireless sensor networks~\cite{BeregK09, DBLP:conf/algosensors/ChanK12,ChanK14, KumarLA05, KormanLSS18} and robot motion planning~\cite{lavalle,hauser2014minimum,KrontirisB17,DBLP:journals/asc/XuM16}. 

Unlike the $s$-$t$-{\sc Shortest Path} problem, \mcp is computationally intractable. It is known to be {\sf NP}-hard~\cite{carr2000red}
as well as hard from the perspective of parameterized algorithms~\cite{EibenK20}.
It is also known that, unless ${\sf P = NP}$, \mcp cannot be approximated within a factor of $O(2^{\log^{1-\delta} n})$, for 
$\delta = 1/\log \log^c n$ and some constant $c < 1/2$, even when the graph is planar~\cite{carr2000red}. In particular,
Kumar~\cite{Kumar19} showed  that, assuming the stronger \dvrconj{} conjecture~\cite{chlamtavc2017minimizing}, there is no 
polynomial time algorithm with approximation factor less than $n^{(\frac{1}{8}-\epsilon)}$, for any fixed $\epsilon >0$. 
There are only some limited results in the positive direction. In the case where the colors are on the \emph{edges} instead of 
vertices and each edge has \emph{only} one color, Hassin et al.~\cite{HassinMS07} presented an $O(\sqrt{n})$-approximation algorithm. 
For the general \mcp in vertex-colored graphs, the best bound known is an $O(\sqrt{n})$ approximation due to 
Bandyapadhyay et al.~\cite{bandyapadhyay2018improved}.

Due to the intractability of \mcp, research has focused on more tractable versions of the problem, and many variants of this problem 
have been explored in different communities.
Perhaps the most studied variant is the {\sc Barrier Resilience} problem~\cite{BeregK09, DBLP:conf/algosensors/ChanK12,ChanK14, KumarLA05, KormanLSS18} in wireless networks, where the task is to connect two given points $s$ and $t$ in the plane by a path while touching 
as few obstacles as possible. Because the obstacles model ranges of wireless sensors, they are typically assumed to be disc-shaped.
It is a well-known open problem whether {\sc Barrier Resilience} is polynomial time solvable when all obstacles are unit 
discs~\cite{KormanLSS18}. However, the problem is known to be \np-hard for many other obstacle shapes, including unit-length line segments, 
rectangles with aspect ratio close to one, or non axis-aligned unit squares~\cite{tseng2011barrier, lavalle, eiben2018improved,KormanLSS18}.
Chan and Kirkpatrick~\cite{ChanK14,DBLP:conf/algosensors/ChanK12} gave a simple constant factor approximation for the unit disc case and left as an open problem whether there exists a constant factor approximation algorithm for the case when obstacles are general discs. 

It is easy to see that {\sc Barrier Resilience} is a special case of \mcp. Consider the plane arrangement of the discs, and 
introduce a vertex for each face (connected region of the plane disjoint from disc boundaries) and a vertex for each 
intersection point of two or more disc boundaries. Next connect by edges all pairs of faces that share a common boundary arc,
and each face vertex with all the intersection points on its boundary. Associate a unique color to each disc, and assign to 
each vertex the colors of its incident discs. We now have a natural correspondence between geometric paths in the plane and 
paths (walks) in the graph such that the colors on the path are precisely the obstacles touched by the geometric path.

In Computational Geometry and robotics, a variant of the \mcp problem has been studied under the name {\sc Minimum Constraint 
Removal}~\cite{lavalle,mcp-fpt,hauser2014minimum,KrontirisB17,DBLP:journals/asc/XuM16}. The goal again is to find a path from
a start point $s$ to a target point $t$ in the presence of obstacles, which are typically modeled as simple closed curves in 
the plane (although higher dimensional variants are also considered)~\cite{hauser2014minimum}. To date, the best approximation 
algorithm for {\sc Minimum Constraint Removal} in the plane is a factor $O(\sqrt{n})$-approximation algorithm by 
Bandyapadhyay et al.~\cite{bandyapadhyay2018improved}, who pose the existence of a better approximation algorithm as an open problem. 

The reduction from {\sc Barrier Resilience} to \mcp also works for  {\sc Minimum Constraint Removal}, with one caveat: 
the number of faces and intersection points in the arrangement of the obstacles should be polynomial in the input size
because that determines the size of the graph for \mcp. For most reasonable obstacle models, such as polygons or low-degree splines,
this is indeed the case.  Starting with {\sc Minimum Constraint Removal} the instances of  \mcp produced by this reduction 
have the following properties: (a) the graph $G$ is planar, and (b) the instances are {\em color connected} (the set of vertices containing any color is connected). Indeed, it is easy to see that for this class of instances \mcp can be reduced back to {\sc Minimum Constraint Removal} in the plane: For every color pick a spanning tree and make an obstacle that traces the outline of the spanning tree in the plane.  Our main result is a constant factor approximation algorithm for planar and color-connected instances of \mcp, resolving the open problems of \cite{bandyapadhyay2018improved,DBLP:conf/algosensors/ChanK12,ChanK14}.
\begin{restatable}{theorem}{algomain}
\label{thm:mainAlgPath}
There exists a polynomial time $O(1)$-approximation algorithm for \mcp on color-connected planar graphs.
\end{restatable}
Theorem~\ref{thm:mainAlgPath} immediately implies a polynomial time $O(1)$-approximation algorithm for {\sc Barrier Resilience} 
and for {\sc Minimum Constraint Removal} when the obstacles are sufficiently well behaved (such as general polygons or splines). 
We complement our results by showing that Theorem~\ref{thm:mainAlgPath} is in some ways the best one can hope for---any attempt at
a significant generalization or improvement runs into a wall of computational intractability. First, while the precise constant 
of our approximation ratio may be improved somewhat, the problem does not admit a polynomial time approximation scheme unless 
${\sf P}= \np$. Indeed, \mcp on color-connected planar graphs is at least as hard to approximate as {\sc Vertex Cover}, even 
when the input graph has treewidth $3$ (see Eiben and Kanj~\cite{EibenK20}). This rules out a better than $2$-approximation 
algorithm assuming the Unique Games Conjecture~\cite{DBLP:journals/jcss/KhotR08}  (and better than $\sqrt{2}$ assuming ${\sf P} 
\neq \np$~\cite{DBLP:conf/focs/KhotMS18}). Further, both color-connectivity {\em and} planarity are necessary to get a constant 
factor approximation. We show in Section~\ref{sec:lb-improved} that \mcp  is hard to approximate within a factor of 
$\max \{ m^{1-\epsilon},~n^{1/4 - \epsilon}\}$ assuming a conjecture on the hardness of {\sc Densest 
Subgraph}~\cite{chlamtavc2017minimizing}. The hardness results hold even for very special classes of instances, namely,
diamond paths. Specifically, $G$ is a {\em diamond path} if it can be obtained by replacing each edge $uv$ of a path 
by any number of degree $2$ vertices adjacent to $u$ and $v$ (See Figure~\ref{fig:lb-vertex-colored}). We prove the following 
hardness result, complementing Theorem~\ref{thm:mainAlgPath}.



\begin{restatable}{theorem}{hardnessmain}
 \label{thm:conn-hardness}
Assuming \dvrconj{} conjecture~{\rm \cite{chlamtavc2017minimizing}}, one cannot approximate \mcp within ratio 
$O(m^{1-\epsilon})$ or $O(n^{1/4 - \epsilon})$ in polynomial time, for any $\epsilon > 0$, where $m$ is the number 
of colors and $n$ is the number of vertices in $G$. The bounds hold even on the following two restricted classes of instances. 
\begin{enumerate}
\setlength{\itemsep}{-2pt}
\item $G$ is a diamond path.
\item $G$ has a vertex $v$ so that $G-v$ is a diamond path and $(G,\sigma)$ is color-connected.
\end{enumerate}
\end{restatable}


Although these lower bounds rule out any significant improvement of Theorem~\ref{thm:mainAlgPath}, we do generalize the result in 
a different way, by designing a constant-factor approximation for a {\sc Min-Color} extension of the classical {\sc Prize 
Collecting Steiner Forest} problem. In this extension, we are given an undirected graph $G$, vertex coloring $\sigma$, 
multiple source-destination connection pairs $(s_1, t_1), \dots, (s_k, t_k)$, and a cost $w_i$ for failing to connect the $i$th pair.
The objective is to minimize total number of colors used in all the paths plus the total cost of all the pairs left disconnected. 
The special case with $w_i = \infty$, forcing all pairs to be connected, is the {\sc Min-Color Steiner Forest} problem. 

\begin{restatable}{theorem}{steinerprize}
\label{thm:mainAlgSteinerPrize}
There exists a polynomial time $O(1)$-approximation algorithm for {\sc Prize Collecting Min-Color Steiner Forest} on color-connected planar graphs.
\end{restatable}

While Theorem~\ref{thm:mainAlgSteinerPrize} implies Theorem~\ref{thm:mainAlgPath}, in the interest of a cleaner presentation we 
will prove Theorem~\ref{thm:mainAlgPath} first, and then point out the (minor) modifications needed for the generalization. 
The {\sc Prize Collecting Min-Color Steiner Forest} problem on 
color-connected planar graphs generalizes classic {\sc Steiner Forest} and {\sc Prize-Collecting Steiner Forest} problems on planar graphs. 
Approximation algorithms for the {\sc Steiner Tree} problem and its several variants, such as {\sc Steiner Forest}, {\sc Group Steiner Tree}, {\sc Prize Collecting Steiner Forest}, 
have been widely studied on general graphs and planar graphs. In fact, just on planar graphs there has been significant amount of work done on these problems to design PTASes for them~\cite{DBLP:journals/talg/HajiaghayiKKN12,DBLP:conf/stoc/BateniDHM16,DBLP:journals/jacm/BateniHM11,DBLP:conf/soda/BateniCEHKM11,DBLP:conf/soda/EisenstatKM12,BorradaileKK07,DBLP:journals/talg/BorradaileKM09,DBLP:journals/talg/DemaineHK14}. We believe that the ease with which Theorem~\ref{thm:mainAlgPath} generalizes to Theorem~\ref{thm:mainAlgSteinerPrize} nicely demonstrates the versatility of our methods.


\paragraph{Our Methods.}  
Given a graph $G$ and a pair of vertices $s$ and $t$, let $d(s,t)$ denote the length (number of edges) of the shortest path between $s$ and $t$ in $G$. An $s$-$t$-{\em separator} is a vertex set $S$ such that $s$ and $t$ are in distinct connected components of $G-S$. Shortest paths in (unweighted) graphs admit the following simple {\sf min-max} theorem: $d(s, t) = |{\cal X}| + 1$, where ${\cal X}$ is a maximum cardinality set of pairwise vertex-disjoint $s$-$t$-separators.
The main combinatorial insight behind our algorithm is that a similar (approximate) min-max theorem can be proved for planar and color-connected instances of \mcp. For any set $C \subseteq [m]$ of colors, we define its \emph{host vertex-set} $V(C) \subseteq V$ to be the set of vertices that contain a color in $C$. That is, $V(C) = \{ v \in V ~|~ \sigma(v) \cap C \neq \emptyset\}$. A set of colors $S \subseteq [m]$ is an $s$-$t$ \emph{color separator} if $V(S)$ is an $s$-$t$-separator (see Figure~\ref{fig:color-sep} for an example). 

Let $\cf$ be the set of all $s$-$t$ color separators in $G$. A {\em fractional packing} of color separators is an assignment $0 \leq y_S \leq 1$ for each color separator $S$ in $G$, such that for every color $c$ we have
$$\sum_{\substack{S \in \cf \\ c \in S}} y_S \leq 1 $$
The {\em value} of a fractional packing is $\sum_{S \in \cf} y_S$.

Observe that a family ${\cal X}$ of $s$-$t$ color separators so that no color appears in two different separators in ${\cal X}$ corresponds to a fractional packing of value $|{\cal X}|$. Since every $s$-$t$ path $\pi$ and $s$-$t$ color separator $S$ satisfy $\sigma(\pi) \cap S \neq \emptyset$, it is easy to see that for every $s$-$t$ path $\pi$ and every fractional packing $\{y_S\}$ of $s$-$t$ color-separators, the number of colors on $\pi$ is at least the value of $\{y_S\}$. The engine behind Theorem~\ref{thm:mainAlgPath} is that the inequality holds in the other direction as well, up to constant factors. 

\begin{theorem}\label{thm:approxminmax}
Let $(G, \sigma)$ be an instance of \mcp, $\opt$ be the minimum number of colors on an $s$-$t$-path, and $\tau$ be the maximum value of a fractional packing of $s$-$t$-color separators. Then $\opt = \Theta(\tau)$.
\end{theorem}
We remark that Theorem~\ref{thm:approxminmax} can easily be seen to be false for the instances constructed in our reductions in Theorem~\ref{thm:conn-hardness}, barring us from generalizing it to instances that are either non-planar or not color-connected.  Theorem~\ref{thm:approxminmax} is the engine behind the algorithm of Theorem~\ref{thm:mainAlgPath}, not in the sense that we first prove Theorem~\ref{thm:approxminmax} and then use it to prove Theorem~\ref{thm:mainAlgPath}, but in the sense that we conjectured Theorem~\ref{thm:approxminmax} as a natural extension of the min-max relation for shortest paths, and that the methods for proving Theorem~\ref{thm:approxminmax} led to a proof of Theorem~\ref{thm:mainAlgPath} as well. 

Finding the maximum value fractional packing of color separators can be seen as the following Linear Program.
We will refer to it as \textsc{Packing-LP}.
\begin{align*}
	\text{maximize} \sum_{S \in \cf} y_S~&~\\
	\text{subject to}\hspace{3em}&~\\
	\sum_{\substack{S \in \cf \\ c \in S}} y_S \leq 1 ~&~~ \text{$\forall$ color $c \in [m]$}
\end{align*}
The proofs of both Theorem~\ref{thm:mainAlgPath} and Theorem~\ref{thm:approxminmax} are by first solving and then rounding the dual of 
\textsc{Packing-LP}, which we will refer to as \lp.
The \lp has a polynomial number of variables (one for every color), but an exponential number of constraints (one per color separator). In particular, for each color $i \in \{1, 2, \dots, m\}$, we associate a variable $0 \leq x_i \leq 1$ that indicates whether or not color $i$ is included in the solution. We then have the following formulation.

\begin{align*}
	\text{minimize} \sum_{i \in [m]} x_i~&~\\
	\text{subject to}\hspace{3em}&~\\
    \sum_{j \in S} x_j \geq 1~&~~ \text{$\forall~s$-$t$ color separators}~S \in \cf
\end{align*}

The strong duality theorem of Linear Programming~\cite{papadimitriou1998combinatorial} implies that $\tau$ is equal to the optimum of $\lp$. An inspection of $\lp$ shows that it deals with fractional {\em hitting sets} for the set of all $s$-$t$ color separators. More formally, a {\em color-hitting-set} is a set $C$ of colors so that for all $s$-$t$ color separators $S \in \cf$, $C \cap S \neq \emptyset$. Clearly integral solutions to $\lp$ correspond to color-hitting sets. It turns out that color-hitting sets are strongly tied to minimum color $s$-$t$ paths: a color set $C$ is a color hitting set if and only if there exists an $s$-$t$ path $\pi$ such that $\sigma(\pi) \subseteq C$ (see Lemma~\ref{lemma:hittingConnectingEquiv} for a short proof).

Thus proving Theorem~\ref{thm:approxminmax} amounts to upper bounding the integrality gap of $\lp$ by a constant. Making such a proof algorithmic, in the sense of making a polynomial time procedure to find an optimal fractional solution to $\lp$, and rounding it to an integral solution with only a constant factor loss in the objective function, is then sufficient to obtain a polynomial time approximation algorithm for \mcp and prove Theorem~\ref{thm:mainAlgPath}.

We use the \emph{color intersection graph}, which has a vertex for every color and an edge between two colors if they are incident to a common face.
The key insight (and the main technical part) of our rounding algorithm is that in the color intersection graph, 
if we consider the shortest path metric where the distance between two adjacent colors is the average of the LP-values of the corresponding $x_c$ variables, then no ball of radius $0.4$ can contain a color separator. With this insight, applying the bounded diameter decomposition of Leighton and Rao~\cite{leighton1999multicommodity} would immediately yield an $O(\log n)$ approximation algorithm for \mcp, and a proof of Theorem~\ref{thm:approxminmax}, but with an $O(\log n)$ gap instead of $O(1)$. To get rid of the $O(\log n)$ factor we observe that the color intersection graph is a region graph over a planar graph (see Section~\ref{sec:LP-rounding} for a definition) and apply the improved bounded diameter decomposition of Lee~\cite{lee2016separators} for such graphs. 

The final hurdle to obtaining a polynomial time approximation algorithm for \mcp is that \lp contains an exponential number of constraints, thus our algorithm can not explicitly construct the LP from the instance and run a solver on it. Instead we will use the ellipsoid method with a separation oracle~\cite{papadimitriou1998combinatorial}. A separation oracle for \lp amounts to the following problem: given as input a planar graph $G$, coloring function $\sigma$, and assignment $x_c$ for every color $c$, to determine whether there exists a color separator $S$ such that $\sum_{c \in S} x_c < 1$. This is equivalent to the problem of finding a minimum weight color separator. This problem (or rather its equivalent formulation as finding a minimum cardinality subset of polygonal obstacles whose union separate two points $s$ and $t$ in the plane) is known as $2$-{\sc Point Separation}, for which a polynomial time algorithm was given quite recently by Cabello and Giannopoulos~\cite{cabello2016complexity}. We were not aware of this algorithm and designed our own polynomial time algorithm for $2$-{\sc Point Separation}, which turns out to be substantially shorter and (in our opinion) simpler.

We believe that Theorem~\ref{thm:approxminmax} together with the methods for rounding algorithm for \lp will find further applications for obstacle removal problems. We do not obtain an explicit bound on the constant in our approximation ratio - in particular it depends on constants hidden in the big-O notation in the region decomposition theorem of Lee~\cite{lee2016separators}. It is likely that the ``correct'' constant for Theorem~\ref{thm:approxminmax}, and therefore also for an approximation algorithm based on \lp is much smaller. For instances that arise from practical applications, rather than worst case instances, the ratio may be smaller still. Therefore it is interesting both to pin down the ``right'' constant factor approximation ratio for \mcp for planar color connected graphs, as well as evaluate the performance of heuristic algorithms for \mcp based on rounding \lp on practical instances.

\paragraph{Organization.} 
Our paper is organized as follows. 
In Section~\ref{sec:roadmap} we introduce some basic definitions and outline a roadmap for our algorithm. 
One of the key insights here is an alternative characterization of \mcp  
in terms of \emph{color separators} and finding a minimum size color set that \emph{hits} 
all color separators. We use this characterization to obtain a linear program for \mcp on
color-connected planar graphs (abbreviated as \textsc{Planar-Conn-MCP}). The next Section~\ref{sec:colproperties} gives some structural properties of color separators that we use both to solve  \lp{} and to round it. In Section~\ref{sec:LP-rounding}, we will show how 
to round a solution for this LP to obtain our approximation algorithm. 
Our rounding scheme makes use of well-known small diameter graph decompositions due to Leighton and Rao~\cite{leighton1999multicommodity}, and 
Klein, Plotkin and Rao~\cite{klein1993excluded}.  Section~\ref{sec:well-behaved-separators} gives 
proof of Lemma~\ref{lemma:well-behaved} which establishes relationship between color separator  and the corresponding  separating cycle in the dual. 
Section~\ref{sec:computing-color-separators} gives an alternate polynomial time algorithm for $2$-{\sc Point Separation}, which is used as for separation oracle. 
Finally, we discuss some improved hardness of approximations bounds in Section~\ref{sec:lb-improved}.
These bounds are based on the so-called \dvrconj conjecture~\cite{chlamtavc2017minimizing}. We conclude the paper with some interesting open problems in Section~\ref{sec:conclusion}. 

To follow the main algorithmic result of the paper, a reader may omit Sections~\ref{sec:colproperties}, 
and~\ref{sec:well-behaved-separators} in the first pass, and the hardness result of Section~\ref{sec:lb-improved} can 
be read independently.

\begin{section}{Basic Definitions and Roadmap}
\label{sec:roadmap}
In this section, we will introduce the notion of color separators and use them to outline a
roadmap for our algorithm. We begin by setting up some basic definitions.

Let $(G, \sigma)$ denote a \emph{colored graph} that consists of a planar graph $G=(V, E)$
and a coloring function $\sigma : V \rightarrow 2^{[m]}$ that satisfies color-connectivity.
That is, the set of vertices containing any color is connected.
If $\sigma(v) = \emptyset$ for some $v \in V$, we say that $v$ is a \emph{white vertex}.
Without loss of generality, we can assume that both $s$ and $t$ are white vertices.
This holds because every $s$-$t$ path must include all colors from both $s$ and $t$. 
For any set $C \subseteq [m]$ of colors, we define its \emph{host vertex-set} $V(C) \subseteq V$ 
to be the set of vertices that contain a color in $C$. 
That is, $V(C) = \{ v \in V ~|~ \sigma(v) \cap C \neq \emptyset\}$.
Assuming that $s$ and $t$ are connected in $G$, we can now define color separators formally as follows.

\begin{Definition}[Color Separator]
	\label{def:color-sep}
	A set of colors $S \subseteq [m]$ is an $s$-$t$ \emph{color separator} if $s$ and $t$ are disconnected in $G - V(S)$.
	(See Figure~\ref{fig:color-sep} for an example.)
\end{Definition}

\begin{figure}[t!]
\centering
	\includegraphics{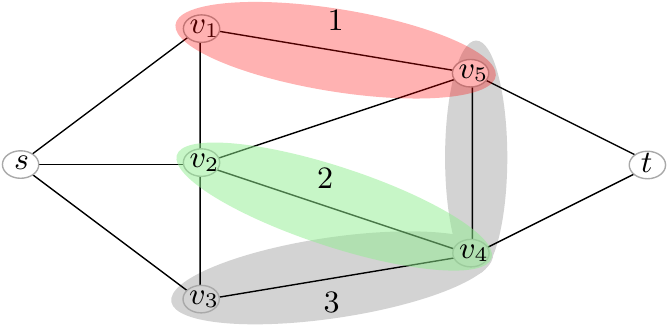}
	\caption{An instance of \textsc{Planar-Conn-MCP} with three colors. 
					 Color sets $\{1,2\}$ and $\{3\}$ are both color separators.}
	\label{fig:color-sep}
\end{figure}

Let $\cf$ to be the set of all $s$-$t$ color separators of $G$, then we say that a set of colors $C$ is a \emph{color-hitting set} if $C \cap S \neq \emptyset$ for all $s$-$t$ color separators $S \in \cf$. We have the following lemma.

\begin{lemma}\label{lemma:hittingConnectingEquiv}
Let $C$ be a set of colors. Then $C$ is a color-hitting set if and only if there exists an $s$-$t$ path $\pi$ such that $\sigma(\pi) \subseteq C$.
\end{lemma}

\begin{proof}
For the reverse direction, consider a path $\pi$ such that $\sigma(\pi) \subseteq C$. Every $s$-$t$ color separator $S$ must intersect $\sigma(\pi)$, because otherwise $\pi$ is an $s$-$t$ path in $G - V(S)$, contradicting that $S$ is a color separator. So $S$ must have non-empty intersection with  $\sigma(\pi)$ and  therefore also with $C$.

Suppose $C$ is a color-hitting set. Define the coloring function $\sigma'$ as follows, for every vertex $v$, $\sigma'(v) = \sigma(v) \setminus C$. We claim that $G$ contains a path $\pi$ such that $\sigma'(\pi) = \emptyset$. Suppose this is not true, that is, every $s$-$t$ path in $(G, \sigma')$ contains at least one color. Then the set of all remaining colors $S' = \bigcup_{v \in V} \sigma'(v)$ is also a color separator. However, we have that $C \cap S' = \emptyset$ which contradicts that $C$ is a color-hitting set.	Therefore, $\sigma'(\pi) = \emptyset$, which gives $\sigma(\pi) \subseteq C$.
\end{proof}

An $s$-$t$ path that uses minimum number of colors is called a {\em min-color path}. 
An immediate corollary of Lemma~\ref{lemma:hittingConnectingEquiv} is that the minimum number of colors on an $s$-$t$ path is the same as the size of the smallest color-hitting set. Specifically, for any color-hitting set $C$, Lemma~\ref{lemma:hittingConnectingEquiv} gives a path $\pi$ with $\sigma(\pi) \subseteq C$.
Similarly, for any $s$-$t$ path $\pi$, the set $C = \sigma(\pi)$ is a color-hitting set.

\begin{lemma}
	\label{lemma:mcp-mhs} let $\pi^*$ be a min-color path, and let $C^*$ be the smallest color-hitting set, then $|\sigma(\pi^*)| = |C^*|$. 
\end{lemma}

The above lemma shows that computing a min-color path is equivalent to computing the color-hitting set of smallest cardinality.
 In the following, we use this equivalence to obtain a linear program for min-color path.

\paragraph{An LP Formulation.} For each color $i \in \{1, 2, \dots, m\}$, we associate a 
variable $0 \leq x_i \leq 1$ that indicates whether or not color $i$ is included in the solution. 
We then have the following formulation which we will refer to as \lp.

\begin{align}
	\text{minimize} \sum_{i \in [m]} x_i~&~ \nonumber \\
	\text{subject to}\hspace{3em}&~\nonumber \\
    \sum_{j \in S} x_j \geq 1~&~~ \text{$\forall~s$-$t$ color separators}~S \in \cf \label{eqn:LP}
\end{align}

It is easy to see that an integral solution to this LP is a color-hitting set of smallest cardinality.
In order to obtain an approximation algorithm using this LP, we first need to be able to solve it
in polynomial time. Although the LP may contain an exponential number of constraints (as the number of
color separators can be exponential), we can solve it in polynomial time using the ellipsoid method provided the existence of a 
polynomial time \emph{separation oracle}. We use a weighted version of color separators to define
a \emph{min-color separator} and present a polytime algorithm for computing it, which will serve as 
the separation oracle for \lp.

\begin{Definition} [Min-Color Separator]~~
\label{def:min-color-separator}
Suppose each color $i \in [m]$ has a non-negative weight $w_i$ associated with it.
Then, a \emph{min-color separator} of $G$ is a color separator 
$S \subseteq [m]$ that minimizes the weight $w(S) = \sum_{i \in S} w_i$.
\end{Definition}

In Section~\ref{sec:computing-color-separators}, we will present a polynomial time algorithm for
computing a min-color separator. Here we state the resulting lemma.
\begin{lemma}
	\label{lemma:min-color-sep}
	A min-color separator can be computed in polynomial time on planar graphs with color connectivity.
\end{lemma}
We note that the problem of computing a min-color separator is essentially the same as the \textsc{2-Point Separation} problem studied by Cabello and Giannopoulos~\cite{cabello2016complexity}, and therefore we could also use the polynomial time algorithm from~\cite{cabello2016complexity} for the separation oracle. However, we did not know about that result and independently obtained an alternative polynomial time algorithm for min-color separator, which is arguably much simpler. Using either of these algorithms as the separation oracle, we obtain the next lemma.

\begin{lemma} \label{lemma:sep-oracle-mcp}
	The \lp can be solved in polynomial time on planar graphs with color-connectivity.
\end{lemma}

The key challenge now is to \emph{round} the fractional solution obtained by solving \lp{}
to obtain an $O(1)$-approximation. We do this in the next two sections. 

\begin{section}{Properties of Color Separators}
\label{sec:colproperties}
In this section, we will discuss some structural properties of color separators that we will use both to solve ~\lp{} and to round it.  It is a well known fact~\cite{itai1979maximum} that an $s$-$t$
cut in a planar graph corresponds to a \emph{separating cycle}, which is a cycle in
the dual graph that separates $s$ from $t$. Inspired by this, we will establish an equivalence
between color separators and cycles in the dual graph. 
We begin by fixing an embedding of $G$ and let $ G^* = (V^*, E^*)$ be its dual graph. 
We then have the following lemma.

\begin{lemma}
	\label{lemma:sep-cycle}
	For every color separator $S$ of graph $G$, there exists a non-empty
	family of separating cycles $\Gamma(S)$ in the dual graph $G^*$.
\end{lemma}
\begin{proof}
	Let $E(S) \subseteq E$ be the set of edges adjacent to vertices in $V(S)$.
	Consider any set $E_\gamma \subseteq E(S)$ such that removing $E_\gamma$ 
	from $G$ disconnects $s$ from $t$. That is, edges $E_\gamma$ are \emph{cut edges}.
	Given a set of cut edges, it is not hard to obtain a separating cycle $\gamma$
	in the dual graph: draw a simple closed curve enclosing one of $s$ or $t$
	and only intersecting the cut edges. 
	
	Repeating this for all possible $E_\gamma$ gives a family $\Gamma(S)$ of 
	separating cycles. Note that $\Gamma(S)$ is non-empty because 
	removing $V(S)$ separates $s$ from $t$, so the set $E_\gamma = E(S)$
	is a trivial cut edge set.
\end{proof}

Therefore, for each color separator $S$, we can associate a separating cycle $\gamma \in \Gamma(S)$ in the
dual graph. Next, we assign colors to the vertices of the dual graph so that the colors on vertices 
of $\gamma$ correspond to the colors in $S$.

\paragraph{Coloring the Dual Graph} It will be convenient to first extend the coloring function $\sigma$ 
	to the edges of $G$ and $G^*$. That is, for all $e = (u, v) \in E$, we assign $\sigma(e) = \sigma(u) \cup \sigma(v)$.
	We can then extend it to edges of the dual graph $G^*$ in a natural way.
	That is, $\sigma(e^*) = \sigma(e)$ where $e^*$ is the dual edge of $e$.
	Finally, we extend the coloring to dual vertices as $\sigma(v^*) = \bigcup_{e^* \in ~\textit{adj}(v^*)} \sigma(e^*)$,
	where $\textit{adj}(v^*)$ denotes the set of dual edges adjacent to $v^*$. We note the following.

	\begin{observation}
		\label{obs:dual-face-colors}
		If $v^* \in  V^*$ is a dual vertex and $f$ is its corresponding face in $G$, then $\sigma(v^*) = \bigcup_{v \in \partial f} ~\sigma(v)$.
		Here $\partial f$ is the set of vertices on the boundary of face $f$.
	\end{observation}

 \begin{lemma}
 	\label{lemma:color-connected}
 	The colored dual graph $(G^*, \sigma)$ is color-connected.
 \end{lemma}
 \begin{proof}
 		Let $v^*_i, v^*_j$ be two vertices in the dual (faces of $G$).
		Consider a color $c \in \sigma(v^*_i) \cap \sigma(v^*_j)$. By color-connectivity of $G$, there
		must be a path $\pi = v_1 \rightarrow v_2 \rightarrow \dots \rightarrow v_r$ in $G$ from 
		a vertex $v_1$ on the boundary of face $v_i^*$ to a vertex $v_r$ on boundary of face $v_j^*$. 
		We want to find a color-connected path $\pi^*$ from $v^*_i$ to $v^*_j$ in the dual. 
		
		We do an induction on the number of vertices $r$ on the path $\pi$.
		If $v_2$ also lies on face $v_i^*$, we can drop $v_1$ from $\pi$ and we are done by induction.
		If $v_2$ lies on some other face $v_k^*$, we consider the clockwise order of edges $E_1 = \textit{adj}(v_1)$.
		From our coloring scheme, it follows that all edges in $E_1$ contain color $c$ and therefore, all their
		dual edges $E_1^*$ will also contain color $c$. 
		Since $v_i^*, v^*_k$ are both faces adjacent to $v_1$, we can reach $v^*_k$ from $v_i^*$ by
		traversing edges of $E^*_1$ in clockwise order. Moreover, all the dual vertices and edges in this path 
		contain color $c$. Now, we can consider the subpath of $\pi$ from $v_2$ to $v_r$ and will 
		again be done by induction.
 \end{proof}

 Now that we have added colors to the dual graph $G^*$, we want to assign an ordering
 of colors of the separator $S$ on the cycle $\gamma$. We do this by a \emph{labeling function}
 $\lambda : E^* \rightarrow S$ that assigns to every edge $e^* \in \gamma$, a non-white color
 from $\sigma(e^*) \cap S$. This is always possible due to the following lemma.

 \begin{lemma}
 	\label{lemma:non-empty-labeling}
 	The set $\sigma(e^*) \cap S$ is non-empty for all $e^* \in \gamma$.
 \end{lemma}
 \begin{proof}
 		Let $e = (u, v)$ be the primal edge in $G$ corresponding to $e^*$.
		Since $e$ is a cut edge for the color separator $S$, it must be adjacent to 
		some vertex in $V(S)$. Therefore, the color set $\sigma(e^*) = \sigma(e) = (\sigma(u) ~\cup~ \sigma(v))$ 
		contains at least one color from $S$.
 \end{proof}

 We can now easily extend the labeling function $\lambda$ to vertices of $\gamma$ as follows.
 Let the vertices on $\gamma$ be arranged in clockwise order:
 $v^*_1 \rightarrow v^*_2 \rightarrow \cdots \rightarrow v^*_r \rightarrow v^*_1$.
 If $e_i^* = (v_{i-1}^*, v^*_i)$ is the edge preceding $v^*_i$ in this order,
 We simply assign $\lambda(v_i^*) = \lambda(e_i^*)$.
 Since $\sigma(e_i^*) \subseteq \sigma(v_{i-1}^*) \cap \sigma(v_i^*)$, we have the following.

 \begin{observation}
 	\label{obs:switch-at-face}
 	If $v_{i-1}^*$ and $v_i^*$ are consecutive vertices on $\gamma$, then $\lambda(v_i^*) \in \sigma(v_{i-1}^*)$.
 \end{observation}

 This gives us a cyclic sequence of colors $\lambda(v^*_1) \rightarrow \lambda(v^*_2) ~\dots~ \leadsto ~\dots~ \lambda(v^*_r) \rightarrow \lambda(v^*_1)$ 
 which we will refer to as \emph{color-cycle} of $\gamma$ and denote it by $\lambda(\gamma)$. 
 Intuitively, the labeling function $\lambda$ simply maps a separating cycle $\gamma \in \Gamma(S)$ to a color-cycle $\lambda(\gamma)$ 
 by \emph{selecting} one color belonging to $S$ from every vertex on $\gamma$.
 However, we still need a little more structure on how the colors of $S$ appear on 
 the color-cycle $\lambda(\gamma)$. Towards that end, we establish the notion of 
 a \emph{well-behaved} separating cycle.

\begin{Definition}[Well-behaved Separating Cycle]
We say that a color-cycle $Z$ is \emph{well-behaved} if all occurrences of any given color $c$ 
in $Z$ are consecutive. Then $\gamma$ is a \emph{well-behaved separating cycle} if there
exists a labeling function $\lambda$ such that the color-cycle $\lambda(\gamma)$ is well-behaved.
\end{Definition}

For example, the color-cycle $Z_1 = c_1 \rightarrow c_2 \rightarrow c_2 \rightarrow c_1$ 
is well-behaved, but $Z_2 = c_1 \rightarrow c_2 \rightarrow c_3 \rightarrow c_2 \rightarrow c_1$ is not.
The following lemma states that a well-behaved separating cycle always exists.
For the sake of clarity, we defer its technical details to Section~\ref{sec:well-behaved-separators}.
\begin{lemma}
	\label{lemma:well-behaved}
	For any color separator $S$, there exists a separating cycle $\gamma$ and its labeling $\lambda : V^* \rightarrow S$
	such that the color-cycle $\lambda(\gamma)$ is well-behaved.
\end{lemma}

We also note the following important property of separating cycles that follows from the Jordan curve theorem~\cite{hales2007jordan}.
\begin{lemma}
	\label{lemma:odd-crossings}
	Let $\gamma$ be a simple closed curve in the plane and let $\pi_{st}$ be a simple path between two points $s, t$
	in the plane. Then $\gamma$ \emph{separates} $s$ and $t$ if and only if the number of times $\pi_{st}$ intersects
	$\gamma$ is odd.
\end{lemma}

Combining Lemma~\ref{lemma:well-behaved} and~\ref{lemma:odd-crossings}, we have the following.
\begin{lemma}
	\label{lemma:color-cycle-odd-crossings}
	Let $(G, \sigma)$ be a color-connected planar graph and $(G^*, \sigma)$ be its colored dual graph.
	Moreover, let $\pi_{st}$ be an arbitrary $s$--$t$ path in $G$.
	Then for any color separator $S$ of $(G, \sigma)$, there exists a \emph{well-behaved} cycle $\gamma$ and its labeling $\lambda$
	in $G^*$ such that $\gamma$ crosses $\pi_{st}$ an odd number of times.
\end{lemma}

\end{section}

\begin{section}{\boldmath An $O(1)$-Approximation Algorithm}\label{sec:LP-rounding}
In the previous sections, we introduced the notion of color separators, established some useful properties, and designed a linear program \lp and claimed that it can be solved in polynomial time~(Lemma~\ref{lemma:sep-oracle-mcp}). Recall that a fractional solution of \lp corresponds to a solution vector $\hat{x} = \langle x_1, x_2, \dots, x_m \rangle$ such that $\sum_i x_i \leq \opt$ where $\opt$ is the number of colors used by a min-color path. In this section, our goal is to round $\hat{x}$ to compute an integer solution vector $\hat{y} = \langle y_1, y_2, \dots, y_m\rangle$ such that $\sum_i y_i \leq a_0 \cdot \opt$ for some constant $a_0$. 
Given a colored graph $(G, \sigma)$, let $\cg = (\cc, \ce)$  be its \emph{color-intersection graph} defined as follows. For every color $i \in [m]$, we add a vertex $c_i \in \cc$. Moreover, we add an edge $(c_i, c_j) \in \ce$ if the colors $i$ and $j$ \emph{intersect} at some dual vertex $v^* \in G^*$. That is, $\{i, j\} \subseteq \sigma(v^*)$.

Our rounding scheme is based on the well-known \emph{small diameter graph decomposition} method. (See e.g~\cite{leighton1999multicommodity} and~\cite{klein1993excluded}.) We will need a node-weighted version of this decomposition where distance values are on nodes and distance between two nodes is defined as the sum of the distance values of the nodes on the shortest path between them, and we will be applying the decomposition to the color intersection graph $\cg = (\cc, \ce)$. To state the decomposition result that we need, we start with a few definitions. 

Suppose $G$ is a graph with a distance function $d$ that assigns a non-negative value $d(v)$ to every vertex $v$. We extend the distance function $d$ to the edges of $G$ as follows: the edge $uv$ gets the distance $d(uv) = (d(u) + d(v))/2$. The distance function can now be further extended to pairs of vertices, so that $d(u, v)$ is the shortest path distance between $u$ and $v$ in the edge-weighted graph with edge weights defined by the function $d$. The diameter of a vertex set $X \subseteq V$ is $\max_{u, v \in X} d(u, v)$. Notice that when we look at the diameter of a vertex set $X$ we are looking at shortest path distances between vertices of $X$ in the entire graph $G$, not the shortest path distances in $G[X]$. We will prove the following lemma. 

%
%
\begin{lemma} \label{lemma:small-diamater-colors}
If $\cg = (\cc, \ce)$ is the color-intersection graph of $(G, \sigma)$, and each vertex $c_i \in \cc$ has a distance value of $d(c_i)$, then for every $\Delta > 0$ there exists a set of vertices $X \subseteq \cc$ such that the diameter of each component of $\cg - X$ is at most $\Delta$ and  $|X| = O(1/\Delta) \cdot \sum_{c_i \in \cc} d(c_i)$. Furthermore, such a set $X$ can be computed from $\cg$ and $d$ in polynomial time.
\end{lemma}  

Lemma~\ref{lemma:small-diamater-colors}, but with the weaker bound $|X| = O(\log m/\Delta) \cdot \sum_{c_i \in \cc} d(c_i)$ follows quite directly from the work of Leighton and Rao~\cite{leighton1999multicommodity}. Here $m$ is the number of colors, and this would lead to an $O(\log m)$ approximation for min-color path. To prove Lemma~\ref{lemma:small-diamater-colors} as stated we shall use that $\cg$ is a {\em region intersection graph} over a planar graph, and use a decomposition theorem of Lee for such graphs~\cite{lee2016separators}. We say that a graph $G = (V, E)$ is a \emph{region intersection} graph over the \emph{base} graph $G_0 = (V_0, E_0)$ if for every $u \in V$ there exists a connected set of vertices $R_u \subseteq V_0$ such that $(u, v) \in E$ if and only if $R_u \cap R_v \neq \emptyset$.

\begin{observation}\label{obs:cgrig}
$\cg$ is a region intersection graph over a planar graph $G_0$.
\end{observation}

\begin{proof}
Let $G_0$ be the (planar) graph constructed from $G$ by adding a vertex for every face of $G$ and connecting this vertex to every vertex of $G$ incident to the face. We show that $\cg$ is a region intersection graph over the base graph $G_0$. For every vertex $c_i \in \cc$ make the region $R_i$ which contains all vertices $v$ of $G_0$ that are vertices of $G$ and $i \in \sigma(v)$, and all vertices $v^*$ of $G_0$ that are faces of $G$ and $i \in \sigma(v^*)$. Note that $R_i$ is connected in $G_0$ for every $i$, and that $c_ic_j \in \ce$ if and only if $R_i \cap R_j \neq \emptyset$.
\end{proof}
%
%
For any vertex $c \in V$, define a ball centered at $c$ of radius $R \geq 0$ as
\begin{align*}
	\cb(c, R) = \left\{ v \in V : d(c, v) < R - \frac{d(v)}{2} \right\}
\end{align*}
A distribution over subsets $S \subseteq V$ is an \emph{$(\alpha, \Delta)$-random separator} if (i) for all $v \in V$ and $R \geq 0$, the probability $\mathbf{P}[~\cb(v, R)$ does not intersect $S~] ~\geq~ 1 - \alpha \frac{R}{\Delta}$, and (ii) the diameter of every component of $G - S$ (in the metric $d$) is at most $\Delta$. We now state the main result of Lee~\cite{lee2016separators} relevant to us.
\begin{proposition}[Corollary 4.3 in~\cite{lee2016separators}~]
	\label{thm:rig-separators}
	Let $G = (V, E)$ be a region intersection graph over $G_0$ such that $G_0$ excludes $K_h$ as a minor.
	Then for any distance function $d : V \rightarrow \mathbb{R}_+$, the
	graph $G$ admits an $(\alpha, \Delta)$-random separator with $\alpha = O(h^2)$.
\end{proposition}

Putting the above together we obtain the following lemma.

\begin{lemma}\label{lem:sampleSep}
	If $S$ is sampled from an $(\alpha, \Delta)$-random separator of $G$, then each component of $G-S$ has diameter at most $\Delta$ and the expectation of 
	$|S|$ is $O(\frac{\alpha}{\Delta}) \cdot \sum_v  d(v)$.
\end{lemma}

\begin{proof}
Since $S$ is an $(\alpha, \Delta)$-random separator, each component of $G-S$ has diameter is at most $\Delta$.
For the expectated size of $S$, we have that $\mathbf{P}[~\cb(v, R)$ intersects $S~] \leq \alpha \frac{R}{\Delta}$.
From the way random separators are defined in Lee~\cite{lee2016separators}, the probability that 
$\mathbf{P}[v \in S~] ~\leq~ \alpha \frac{d(v)}{\Delta}$ (Remark~2.11~\cite{lee2016separators}).
Therefore, the expectation of $|S|$ is $O(\frac{\alpha}{\Delta}) \cdot  \sum_v  d(v)$.
\end{proof}

We are now ready to complete the proof of Lemma~\ref{lemma:small-diamater-colors}.

\begin{proof}[Proof of Lemma~\ref{lemma:small-diamater-colors}]
Since $\cg$ is a region intersection graph of a planar graph (by Observation~\ref{obs:cgrig}) and planar graphs are $K_5$-minor-free, we may apply Proposition~\ref{thm:rig-separators} on $\cg$, and obtain an $(\alpha, \Delta)$-random separator with $\alpha = O(5^2) = O(1)$. By Lemma~\ref{lem:sampleSep} if a set $X$ is sampled according to this distribution then each component of $G-X$ has diameter is at most $\Delta$, and the expected size of $X$ is $O(\frac{1}{\Delta}) \cdot \sum_v  d(v)$. This establishes the {\em existence} of the set $X$ as claimed by  Lemma~\ref{lemma:small-diamater-colors}.

For a polynomial time algorithm to compute $X$ observe that the random separator of Proposition~\ref{thm:rig-separators} (see Lee~\cite{lee2016separators}) is based on the well-known iterative chopping scheme of Klein, Plotkin and Rao~\cite{klein1993excluded}  , which can be turned into a polynomial time algorithm by, in each iteration of the algorithm, greedily choosing the ``chop value'' that would minimize the total weight of the vertices added to $X$. We omit the details as this is a standard adaptation of~\cite{lee2016separators} but would require a substantial fraction of the text to be reproduced here. 
\end{proof}


We are now all set to describe our approximation algorithm (See Algorithm~\ref{alg:approx}.) In the algorithm, we will set $\epsilon = 0.1$
First we upper bound the size of the color set returned by the algorithm.

\begin{algorithm}
		~\\
		\textbf{\textsc{Input}}: A colored graph $(G, \sigma)$ that is planar graph and color-connected, $\eps = 0.1$. \\
		\textbf{\textsc{Output}}: Set of colors $\cc^*$ that hits all color separators of $G$
		\begin{enumerate}
			\item Using Lemma~\ref{lemma:sep-oracle-mcp}, solve \lp in polynomial time. Let $\hat{x} = \langle x_1, x_2, \dots, x_m \rangle$
						be the fractional solution vector.
			\item \label{item:preprocess} Include all colors $j$ to the solution $\cc^*$ such that $x_j \geq \eps$.
			\item \label{item:color-graph} Build the \emph{color intersection} graph $\cg = (\cc, \ce)$ over the colors not already included in $\cc^*$.
						Assign distance values to the nodes of $\cg$ as $d(c_j) = x_j$, where $c_j \in \cc$ is the vertex for color $j$.
			\item \label{item:apply-decomp} Apply Lemma~\ref{lemma:small-diamater-colors} on the node-weighted graph $\cg$ with diameter $\Delta = \frac{1}{2} - \eps$.
						Let $X$ be the set of cut vertices obtained from the lemma.
			\item For each $c_j \in X$, add the corresponding color $j$ to the solution $\cc^*$. Return $\cc^*$.
		\end{enumerate}
		\caption{Approximate \textsc{Planar-Connected-MCP}}
		\label{alg:approx}
\end{algorithm}

	\begin{lemma}\label{lem:boundColors} If $\opt$ is the minimum number of colors of an s-t path, then the number of colors $|\cc^*|$ returned by Algorithm~\ref{alg:approx} is at most $O(\opt)$.
	\end{lemma}

	\begin{proof}
		Rounding up variables with $x_j \geq \eps$ in Step~\ref{item:preprocess} of the algorithm only 
		increases the cost by a constant factor of $1/\eps$. Moreover in Step~\ref{item:apply-decomp},
		we apply Lemma~\ref{lemma:small-diamater-colors} with $\Delta = \frac{1}{2} - \eps = 0.4$, which gives $|\cc^*| = O(1/\Delta) \cdot \opt ~=~ O(\opt)$. 
	\end{proof}
	
	In the next two lemmas, we show that Algorithm~\ref{alg:approx} indeed computes a set of colors $\cc^*$ that 
	hits all color separators of $G$. For the ease of exposition, we will implicitly use a set of colors to also 
	refer to the set of corresponding vertices of graph $\cg$.

	\begin{lemma}
		\label{lemma:connected-in-GI}
		If $S$ is an inclusion minimal color separator, then the colors in $S$ are connected in $\cg$.
	\end{lemma}
	\begin{proof}
		From Lemma~\ref{lemma:well-behaved}, we know that for every color separator there exists a well-behaved separating cycle $\gamma$ in $G^*$ and its labeling $\lambda : V^* \rightarrow S$. Since $S$ is inclusion minimal, this labeling is surjective. Consider any two consecutive vertices $v^*_{i-1}, v^*_{i} \in \gamma$ that get a different label. That is, $\lambda (v^*_{i-1}) = j$, $\lambda (v^*_{i}) = k$ and $j \neq k$. From Observation~\ref{obs:switch-at-face} it follows that $k \in \sigma(v^*_{i-1})$ and therefore $(c_j, c_k)$ must be an edge in $\cg$. Since $\lambda(\gamma)$ is a well-behaved color-cycle, the colors in $S$ form a cycle in $\cg$.
	\end{proof}

	\begin{lemma}
		\label{lemma:tree-shortcut}
		The set of colors $\cc^*$ returned by Algorithm~\ref{alg:approx} hits all $s$-$t$ color separators.
	\end{lemma}
	\begin{proof}
		The proof is by contradiction. Suppose $S$ is an (inclusion minimal) color separator that is not hit by $\cc^*$.
		Then, $S$ must be contained in a single component $\kappa'$ of $\cg - \cc^*$. 	This holds because colors in $S$ are connected in $\cg$ (Lemma~\ref{lemma:connected-in-GI}) and can be split into different components only if $S \cap \cc^* \neq \emptyset$. 
		Let $c_1$ be a color in $S$ and let $\kappa$ be the set of colors in a ball in $\cg{}$ of radius $1/2-\epsilon$ around $c_1$. 
		Observe that $\kappa' \subseteq \kappa$ by the choice of $\Delta$.
		
We will now focus only on colors that lie in $\kappa$. First, we sort and rename the colors in $\kappa$ as $\{c_1, c_2, \dots, c_r\}$ by their distance from 
the central color $c_1$ of the ball $\kappa$. When renaming, fix a shortest path tree $T$ of $\kappa$ rooted at $c_1$ and break ties so that the predecessor $c_j$ of every color $c_i$ (with $i > 1$) in $T$ satisfies $j < i$.
Next, we define $C_\ell = \{c_1, c_2, \dots, c_\ell\}$ to be the color set containing the first $\ell$ colors.
		
		Observe that $C_1$ is not a color separator because $c_1 \notin \cc^*$ and therefore in \lp{} we have that the corresponding variable $x_1$ is strictly less than $1$. 
		 If $C_1 = \{c_1\}$ had been a color separator the constraints of  \lp would have forced $x_1 = 1$. Further, observe that $C_r$ contains $S$ and therefore $C_r$ is a color separator. Let $j > 1$ be the smallest index such that the set $C_j$ is a color separator and $C_{j-1}$ is not a color separator. 
		
		Since $C_{j-1}$ is not a color separator, we can find an $s$--$t$ path $\pi_{st}$ in the \emph{primal} graph $G$ disjoint from $V(C_{j-1})$. That is, no vertex on $\pi_{st}$ contains a color from $C_{j-1}$.	From the way we assign colors to the dual graph, it follows that no edge on $\pi_{st}$ contains a color from $C_{j-1}$ 	and therefore every dual edge that crosses $\pi_{st}$ does not contain a color from $C_{j-1}$. This proves the following claim.

\begin{claim}\label{claim:edge-is-disjoint}
		Every dual edge that contains a color from $C_{j-1}$ does not cross $\pi_{st}$.
\end{claim}
		
		\begin{figure}[htb!]
		\centering
			\includegraphics{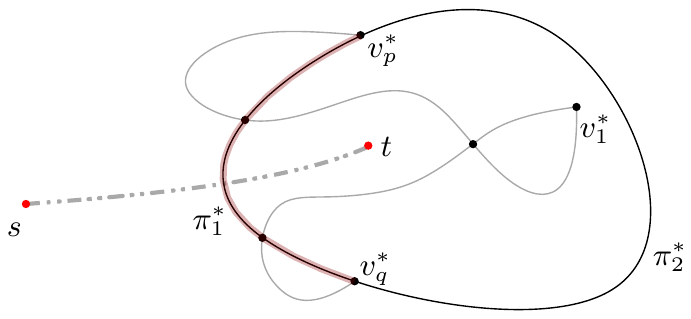}
			\caption{Modifying $\gamma$ to obtain another separating cycle $\gamma'$ of weight strictly less than $1$.
							The primal path $\pi_{st}$ is shown in dash-dotted. Only edges of $\pi_1^*$ (shown highlighted) 
							can cross $\pi_{st}$.}
		\label{fig:no-separator-in-ball}
		\end{figure}

Now consider the separating cycle $\gamma \in G^*$ for the color separator $C_j$ and its well behaved labeling $\lambda$ that 
maps every edge of $\gamma$ to a color in $C_j$. 
%
	%
		Let $\pi^*_1 = v_p^* \leadsto v_q^*$ be the maximal subpath of $\gamma$ such that $\lambda(e^*) = c_j$ for all $e^* \in \pi^*_1$.
		Let $\pi^*_2 = \gamma - \pi^*_1$ be the remainder of $\gamma$.
		Since all occurrences of $c_j$ are consecutive, we have that $\lambda(e^*) \in C_{j-1}$
		for all $e^* \in \pi^*_2$. 	Therefore, from Claim~\ref{claim:edge-is-disjoint}, it follows that the path $\pi^*_2$ does not cross $\pi_{st}$.
		Since $\gamma$ crosses $\pi_{st}$ an odd number of times, we obtain the following.
		\begin{claim}
			\label{claim:crossing}
			The path $\pi_1^*$ crosses $\pi_{st}$ an odd number of times.
		\end{claim}
		   
		Observe that vertices $v_p^*$ and $v_q^*$ both contain at least one color in $C_{j-1}$.
		This holds because they are both adjacent to an edge $e^* \in \pi_2^*$ such that $\lambda(e^*) \in C_{j-1}$.
		Let $c_p, c_q$ be two colors from $C_{j-1}$ that lie on vertices $v^*_p$ and $v^*_q$ respectively.
		Moreover, let $v_1^*$ be any vertex in the dual graph that contains color $c_1$.
		
		Let $\hat{C}_p$ be the set of colors on the path from $c_p$ to $c_1$ in $T$, and let $\pi_p^* = v_1^* \leadsto v^*_p$ be a path from $v^*_1$ to $v^*_p$ 
        with a corresponding labeling $\lambda'$ that assigns to every edge $e^*$ in $\pi_p^*$ a color from $\hat{C}_p \cap \sigma(e^*)$.
%
%
%
Since the radius of $\kappa$ is at most $1/2 - \eps$, the sum of distance values of colors in $\hat{C}_p$ is also at most $1/2 - \eps$. Moreover, $\hat{C}_p \subseteq C_{j-1}$ since the predecessor of every vertex in $T$ is eariler in the ordering $c_1, \ldots, c_r$. Thus, every edge in $\pi_p^*$ contains a color from $C_{j-1}$, and therefore it follows from Claim~\ref{claim:edge-is-disjoint} 
that $\pi_p^*$ does not cross $\pi_{st}$. By a symmetric argument, there also exists a similar path $\pi_q^* : v_1^* \leadsto v^*_q$. (See also Figure~\ref{fig:no-separator-in-ball}.) Thus we have proved the following claim. 

		\begin{claim}
			\label{claim:no-crossing}
			The paths $\pi_p^*$ and $\pi^*_q$ do not cross $\pi_{st}$.
		\end{claim}
	
		Next, we combine these paths to obtain a closed walk: $W^* = v_1^* \xrightarrow{\pi^*_p} v^*_p \xrightarrow{\pi_1^*}  v^*_q \xrightarrow{\pi^*_q} v^*_1$
		in the dual graph. The walk simply picks colors given by the labeling $\lambda'$ along the paths $\pi^*_p$ and $\pi^*_q$, 
		so the cost of those colors is $2 \times (1/2 - \eps)$. For the subpath $v^*_p \leadsto v^*_q$ of $\pi^*_1$, 
		the walk picks the color $c_j$ given by labeling $\lambda$. Therefore, the total cost of $W^*$ is $1 - 2\eps  + d(c_j) \leq 1 -2\eps + \eps < 1 $. 
		Note that $d(c_j) < \eps$ because colors with $d(c_j) > \eps$ have already been included into $\cc^*$ during Step~\ref{item:preprocess} of
		the algorithm. 

By Claims~\ref{claim:crossing} and~\ref{claim:no-crossing} the walk $W^*$ crosses $\pi_{st}$ an odd number of times. Among all closed sub-walks of $W^*$ that cross $\pi_{st}$ an odd number of times, pick an inclusion minimal one (using the same labeling). 
This is a simple cycle $\gamma'$ that crosses $\pi_{st}$ an odd number of times, so (by Lemma~\ref{lemma:odd-crossings}) 
the set of $\lambda'$-colors on $\gamma'$ is a color separator.
The cost of the colors on this separating cycle is upperbounded by the cost of $W^*$.
This gives us a color separator $S'$ such that $\sum_{j \in S'} x_j ~<~ 1$, which contradicts the constraint for $S'$ in \lp.
\end{proof}

%

	Lemmata~\ref{lem:boundColors},~\ref{lemma:tree-shortcut},~\ref{lemma:hittingConnectingEquiv},~\ref{lemma:small-diamater-colors} and~\ref{lemma:sep-oracle-mcp} imply that Algorithm~\ref{alg:approx} is a constant factor approximation for min color path on planar, color connected instances. More concretely Lemma~\ref{lemma:sep-oracle-mcp} together with Lemma~\ref{lemma:small-diamater-colors} prove that Algorithm~\ref{alg:approx} runs in polynomial time, Lemma~\ref{lem:boundColors} shows that the size of the set of colors returned by the algorithm is $O(\opt)$, while Lemmata~\ref{lemma:tree-shortcut} and~\ref{lemma:hittingConnectingEquiv} show that there is a path from $s$ to $t$ using only colors from the set $\cc^*$ returned by the algorithm. This proves our main theorem. 
	
\algomain*

The proof of Theorem~\ref{thm:approxminmax} is nearly identical: the proof of Lemma~\ref{lem:boundColors} shows that the set of colors returned by the algorithm is $O(\opt_{LP})$, where $\opt_{LP}$ is the value of the optimum solution to \lp. LP-duality yields that $\opt_{LP} = \tau$, while Lemmata~\ref{lemma:tree-shortcut} and~\ref{lemma:hittingConnectingEquiv} show that there is a path from $s$ to $t$ using only colors from the set $\cc^*$ returned by the algorithm, completing the proof of Theorem~\ref{thm:approxminmax}.


The techniques used to prove Theorem~\ref{thm:mainAlgPath} easily extend to the more general {\sc Min Color Steiner Forest} setting. Here the input is a graph $(G, \sigma)$ and multiple source-destination pairs $(s_1, t_1), \dots, (s_k, t_k)$. The goal is to find a set of paths $\Pi = \{\pi_1, \dots, \pi_k\}$ such that $\pi_i$ is a path connecting $s_i$ to $t_i$ and the total number of colors used by all paths $|\sigma(\Pi)|$ is minimized. Here, $\sigma(\Pi) = \bigcup_{i \in \{1, \dots, k\}} \sigma(\pi_i)$. 

In particular we extend the \lp{} by including in the family ${\cal F}$ the set of $(s_i, t_i)$ separators for all $i$. Observe now that we get a separation oracle for this extended linear program by running the separation oracle of Lemma~\ref{lemma:min-color-sep} for each $s_i$-$t_i$ pair individually. If any of them returns an unsatisfied constraint this is an unsatisfied constraint for the extended LP, because this corresponds to an $s_i-t_i$ color separator $S$ such that $\sum_{i \in S} x_i < 1$.

We now run Algorithm~\ref{thm:mainAlgPath}, but with this extended LP instead of the original one. Observe that (the proof of) Lemma~\ref{lem:boundColors} bounds the number of output colors $|\cc^*|$ in terms of the LP-optimum, not just the optimum for minimum color path. Thus the number of colors returned by Algorithm~\ref{thm:mainAlgPath} is still $O(\opt)$ where $\opt$ is interpreted as the minimum number of colors in a solution to the {\sc Min Color Steiner Forest} instance. 

Finally, Lemma~\ref{lemma:tree-shortcut} never uses that $x_1, \ldots, x_m$ is an {\em optimal} solution to (the original) \lp, only that it is a feasible solution. Thus, because a feasible solution to the extended linear program is a feasible solution to the original \lp simultaneously for all input $(s_i, t_i)$ pairs, Lemma~\ref{lemma:tree-shortcut} yields that the output set $\cc^*$ of colors hits all $(s_i, t_i)$-color separators for every $i$. Lemma~\ref{lemma:hittingConnectingEquiv} then yields that for every $i$ there is a path connecting $s_i$ and $t_i$ using only colors from $\cc^*$. Thus we obtain a constant factor approximation for {\sc Min-Color Steiner Forest}.

\begin{lemma}\label{thm:mainAlgSteiner}
There exists a polynomial time $O(1)$-approximation algorithm for {\sc Min-Color Steiner Forest} on color-connected planar graphs.
\end{lemma}

In fact, the algorithm of Lemma~\ref{thm:mainAlgSteiner} can easily be generalized to a ``Prize-Collecting'' version of the problem. Here every input $(s_i, t_i)$ pair comes with a cost $w_i$, and for every $i$ we have to either connect the pair $(s_i, t_i)$ or pay the cost $w_i$. The objective is to minimize total number of colors used in all of the paths plus the total cost of all the pairs that are left disconnected. We will call this version of the problem {\sc Prize Collecting Min-Color Steiner Forest}.

To lift the algorithm of Lemma~\ref{thm:mainAlgSteiner} to this variant we further extend the \lp by including a variable $0 \leq y_i \leq 1$ for every pair, minimizing $\sum_{i \leq m}x_i + \sum_{i \leq k}y_i \cdot w_i$ and adding $y_i$ on the left hand side of each constraint for $s_i-t_i$ separators. 

The rounding algorithm chooses not to connect the pairs $(s_i, t_i)$ with $y_i \geq 1/2$. For the remaining pairs the variable assignment $x_i' = 2 \cdot x_i$ is a feasible assignment to the linear program for {\sc Min-Color Steiner Forest}, and we may round this solution with total cost $O(\opt)$ using the algorithm from Lemma~\ref{thm:mainAlgSteiner}. This brings us to the most general form of our algorithmic result.

\steinerprize*


\end{section}


\begin{section}{Well-Behaved Color Separators}
\label{sec:well-behaved-separators}
In this section, we present a proof of Lemma~\ref{lemma:well-behaved}. That is, we show that for any 
color separator $S$, there exists a separating cycle $\gamma$ and its labeling $\lambda$ such that
its color-cycle $\lambda(\gamma)$ is well-behaved (all occurrences of any given color are consecutive).

Let $u^*, v^*$ be two non-consecutive vertices on $\gamma$ such that $\lambda(u^*) = \lambda(v^*) = c$.
Suppose we split $\gamma$ at $u^*, v^*$ into two disjoint subpaths $\pi_1 = u^* \leadsto v^*$ 
and $\pi_2 = v^* \leadsto u^*$. That is, $\gamma = \pi_1 \oplus \pi_2$ where $\oplus$ denotes 
the operation of concatenating two paths at their common endpoints.
By color connectivity of $G^*$ (Lemma~\ref{lemma:color-connected}), there must be a simple path $\pi_c$ 
from $u^*$ to $v^*$ (called \emph{shortcut-path}) such that all intermediate vertices on this 
path also contain color $c$. We now have the following lemma.

\begin{lemma}
	\label{lemma:shortcut-colors}
	If $\pi_c$ is \emph{internally disjoint} from $\gamma$, then exactly one of $\gamma_1 =  \pi_1 \oplus \pi_c$ 
	and  $\gamma_2 =  \pi_2 \oplus \pi_c$ is also a separating cycle. 
\end{lemma}

\begin{proof}

\begin{figure}[htb!]
\centering
	\includegraphics{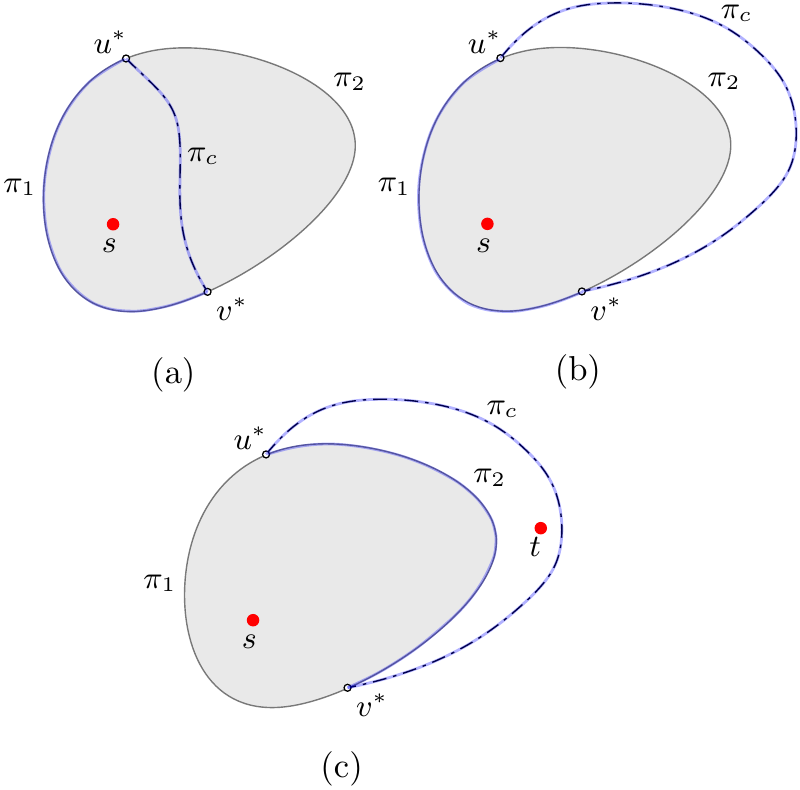}
	\caption{Cases from proof of Lemma~\ref{lemma:shortcut-colors}. Separating cycle $\gamma$ bounds the region $R$ (shown in gray) 
		and is split into paths $\pi_1$ and $\pi_2$ by the shortcut path $\pi_c$. The new separating cycles are shown in bold.}
	\label{fig:shortcut-disjoint}
\end{figure}

	Since $\pi_c$ is internally disjoint from $\gamma$, both $\gamma_1$ and $\gamma_2$ are simple cycles.
	Let $R$ be the region enclosed by $\gamma$, we have the following two cases. We assume that source $s$
	lies inside $R$ and destination $t$ lies outside $R$.
	\begin{enumerate}
		\item \emph{Path $\pi_c$ lies inside $R$}:~~ In this case, $\pi_c$ splits the region $R$ into two disjoint sub-regions
									  defined by $\gamma_1$ and $\gamma_2$. We set $\gamma'$ to whichever of $\gamma_1$ or $\gamma_2$
										contains the source $s$. (See also Figure~\ref{fig:shortcut-disjoint}a.)

		\item \emph{Path $\pi_c$ lies outside $R$}:~~ In this case, one of the two cycles $\gamma_1$ and $\gamma_2$ 
										completely encloses the region $R$ and the other is disjoint from $R$.
										Without loss of generality assume that $\gamma_1$ encloses $R$ and $\gamma_2$
										is disjoint from $R$. 
										If $t$ lies outside $\gamma_1$, we set $\gamma'$ to $\gamma_1$ as our separating cycle as shown in 
										Figure~\ref{fig:shortcut-disjoint}b.
										If $t$ lies inside $\gamma_1$, we set $\gamma'$ to $\gamma_2$ as shown in
										Figure~\ref{fig:shortcut-disjoint}c.
	\end{enumerate}
	In both these cases, we were able to find another separating cycle $\gamma'$, which is basically $\pi_c$ concatenated
	with either $\pi_1$ or $\pi_2$.
\end{proof}

For any color-cycle $Z = \lambda(\gamma)$, we can define the measure $\mu(\gamma, \lambda)$ to be the
number of times $Z$ switches colors. That is, $\mu(\gamma, \lambda)$ is  the number of indices $i$ where $Z_i \neq Z_{i+1}$.
Moreover, let $u^*, v^* \in \gamma$ be two non-consecutive vertices (called a \emph{violating pair}) 
such that $\lambda(u^*) = \lambda(v^*) = c$ and at least one intermediate
vertex on both subpaths $\pi_1, \pi_2$ of $\gamma$ gets a different label.
We then have the following lemma.

\begin{lemma}
	\label{lemma:measure-decreases}
	Let $\gamma$ be a separating cycle and $\lambda$ be its labeling.
	If $u^*, v^*$ is a violating pair of vertices of $\gamma$ and $\pi_c$ is a shortcut-path connecting them,
	then there exists another cycle $\gamma'$ and its labeling $\lambda'$ such that
	$\mu(\lambda', \gamma')$ is strictly less than $\mu(\lambda, \gamma)$.
\end{lemma}

\begin{proof}
	The proof is by induction on the number of vertices in $\gamma \cap \pi_c$.
	The base case is when $|\gamma \cap \pi_c| = 2$. That is, when $\pi_c$ and $\gamma$ are internally disjoint.
	Then using Lemma~\ref{lemma:shortcut-colors} either $\gamma' = \pi_1 \oplus \pi_c$ or $\gamma' = \pi_2 \oplus \pi_c$. 
	In both cases, we can label $\lambda'(w^*) = c$ for all vertices $w^* \in \pi_c$.
	It is easy to verify that $\mu(\lambda', \gamma') < \mu(\lambda, \gamma)$.

	For the inductive step, let $u_1^* \in \gamma \cap \pi_c$ be the vertex closest to $u^*$ on the path $\pi_c$.
	We have the following cases.
	\begin{enumerate}
		\item \emph{$u^*$ and $u_1^*$ are consecutive on $\gamma$}. In this case, we simply modify $\lambda(u_1^*) = c$ which 
					does not increases $\mu(\gamma, \lambda)$. Now we apply induction with $u^* = u^*_1$ and the shortcut $\pi_c$ to be the
					subpath of $\pi_c$ from $u_1^*$ to $v^*$. Clearly, the number of vertices in $\gamma \cap \pi_c$ is one less and by induction
					there must exist another cycle $\gamma'$ and its labeling $\lambda'$ such that
					$\mu(\gamma', \lambda') < \mu(\gamma, \lambda)$.

		\item Otherwise, $u^*$ and $u_1^*$ are non-consecutive on $\gamma$ and the subpath $\pi'_c : u^* \leadsto u_1^*$ of $\pi_c$ is 
					internally disjoint from $\gamma$. The vertices $u^*$ and $u_1^*$ split $\gamma$ into two paths: let
					$\pi_1'$ be the path containing vertex $v^*$ and $\pi_2'$ be the other path. (See also Figure~\ref{fig:shortcut-crossing}a.)
					Next, we apply Lemma~\ref{lemma:shortcut-colors} again with the shortcut-path $\pi_c'$ connecting vertices $u^*, u^*_1$. 
					This gives that either $\gamma_1 = \pi_1' \oplus \pi_c'$ or $\gamma_2 = \pi_2' \oplus \pi_c'$ as a separating cycle.
					In both cases, we modify the label $\lambda(w^*) = c$ for all vertices $w^* \in \pi'_c$.
					That is, the $\pi_c'$ part of the separating cycle does not switch colors.
					We now analyze the number of color switches in both cases as follows.
						
					\begin{figure}[htb!]
						\centering
						\includegraphics{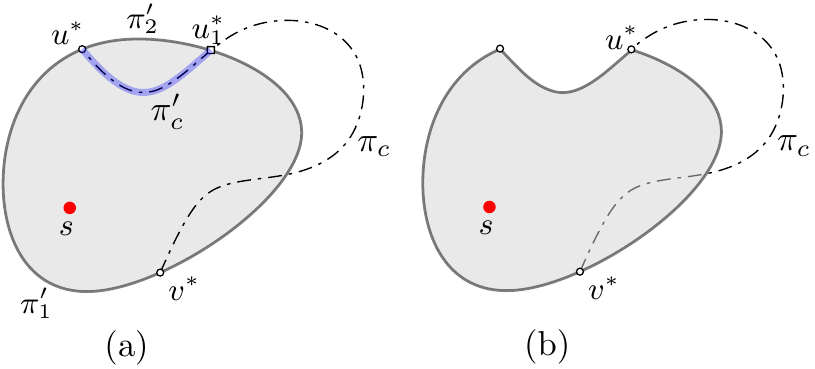}
						\caption{ (a)~ The cycle $\gamma$ with $|\gamma \cap \pi_c| = 4$. Here, $\gamma_1 = \pi_1' \oplus \pi'_c$ is also a separating cycle.
										  (b)~ The cycle $\gamma_1$ with $|\gamma_1 \cap \pi_c| = 3$. The shortcut-path $\pi_c$ is shown dash-dotted.} 
						\label{fig:shortcut-crossing}
					\end{figure}
									
					\begin{enumerate}
						\item \emph{$\gamma_1 = \pi_1' \oplus \pi_c'$ is a separating cycle.} Observe that $\gamma_1$ is obtained by
										replacing the part $\pi_2'$ of $\gamma$ with path $\pi_c'$.
										It is easy to verify that that $\mu(\gamma_1, \lambda) \leq \mu(\gamma, \lambda)$.
										Next we apply induction with $u^* = u^*_1$ and the shortcut-path $\pi_c$ to be the subpath of $\pi_c$ from $u_1^*$ to $v^*$.
										The number of vertices in $\gamma_1 \cap \pi_c$ is one less and by induction there
										must exist another cycle $\gamma'$ and its labeling $\lambda'$ such that 
										$\mu(\gamma', \lambda') < \mu(\gamma_1, \lambda) \leq \mu(\gamma, \lambda)$.
										(See also Figure~\ref{fig:shortcut-crossing}b.)
						
						\item \emph{$\gamma_2 = \pi_2' \oplus \pi_c'$ is a separating cycle.} In this case, observe that $\gamma_2$ is obtained by
						                    replacing the part $\pi_1'$ of $\gamma$ (that contains $v^*$) with the path $\pi_c'$. Indeed, replacing
																$\pi_1'$ gets rid of at least one color switch due to vertex $v^*$ and adds no new color switches.
																Therefore, $\mu(\gamma_1, \lambda) < \mu(\gamma, \lambda)$.
					\end{enumerate}
	\end{enumerate}
\end{proof}

\begin{lemma}
	\label{lemma:well-behaved-restated}
	Every color separator $S$ has a separating cycle that is well-behaved.
\end{lemma}
\begin{proof}
	Let $\gamma \in \Gamma(S)$ be any separating cycle and $\lambda$ be an arbitrary labeling function.
	(From Lemma~\ref{lemma:non-empty-labeling}, such a labeling always exists.)
	Now, if the color-cycle $\lambda(\gamma)$ is well-behaved, we are done.
	If not, there must be a violating pair of vertices $u^*, v^* \in \gamma$ 
	such that $\lambda(u^*) = \lambda(v^*) = c$.
	Therefore we can apply Lemma~\ref{lemma:measure-decreases}, and obtain 
	a separating cycle $\gamma'$ with at least one less color switch.
	Since the number of color switches $\mu(\gamma, \lambda) > 0$, exhaustively
	applying Lemma~\ref{lemma:measure-decreases} gives a separating cycle with 
	no violating pair of vertices.
\end{proof}

\end{section}

\section{Computing a Min-Color Separator}
\label{sec:computing-color-separators}
In this section, we will discuss a polynomial time algorithm to compute the
minimum weight color separator. Specifically, we are given the colored graph $(G,\sigma)$ 
such that each color $j \in [m]$ has a weight $w_j$ associated with it and the goal is to compute 
a color separator $S$ that minimizes  $w(S) = \sum_{j \in S} \wt_j$.

The key to an algorithm for min-color separator is Lemma~\ref{lemma:color-cycle-odd-crossings},
which states that for every color separator $S$, there exists a well-behaved cycle in the colored 
dual graph $(G^*, \sigma)$ that crosses any arbitrary $s$-$t$ path $\pi_{st}$ an odd number of times.
Recall that the notion of well-behaved means that all occurrences of a given color on the cycle 
are consecutive. This lets us formulate the problem of finding a min-color separator as
a shortest path problem in an auxiliary \emph{layered graph} $H$.

\subsection{\boldmath Constructing the Auxiliary Graph $H$}
The input to our construction is the colored dual graph $(G^*, \sigma)$ and an arbitrary $s$-$t$ path $\pi_{st}$.
Roughly speaking, the auxiliary graph $H$ consists of two layers: $L_a$ and $L_b$, with
an identical set of vertices and two types of edges: \emph{intra-layer} edges that go within the layer and \emph{inter-layer} edges 
that go between layers. The graph $H$ will be \emph{edge weighted}. 
We first add vertices and edges to $H$ and later assign weights to its edges.

\paragraph{\boldmath Adding vertices to $H$} 
For every dual vertex $v_i^* \in V^*$, we create $r = |\sigma(v_i^*)|$ copies in $L_a$ and $L_b$, 
one for each color in $\sigma(v^*)$.
More precisely, for every pair $(i, j)$ such that $v_i^* \in V^*$ and $j \in \sigma(v_i^*)$,
we add the vertex $a^j_i$ in layer $L_a$ and $b^j_i$ in layer $L_b$.

Since $G^*$ is a planar graph, we can think of following visualization of $H$ in three dimensions.
Consider $L_a$ to be the bottom layer, $L_b$ to be the top layer, and stack all copies $a_i^j, b_i^j$ of
$v_i^*$ one above another, such that all $a_i^j$ copies come first followed by $b_i^j$. 
(See also Figure~\ref{fig:layered-graph})

\begin{figure}[htb!]
\centering
	\includegraphics{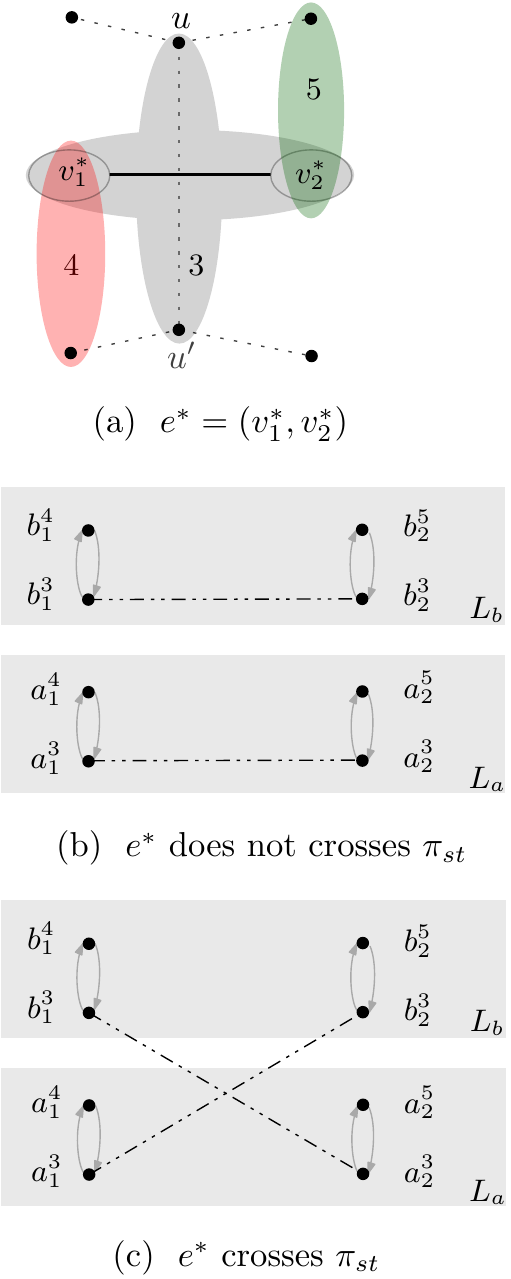}
	\caption{An example of layered graph construction with $\wt_j = 1$ for all colors $j \in [m]$.
					 An edge $e^* = (v_1^*, v_2^*)$ and two cases for adding corresponding edges to $H$ are shown.~
					 In Figure~(b, c),  \emph{free edges} of $H$ are dash-dotted and have weight zero. 
					The \emph{clique edges} of $H$ are shown in gray have weight 1.}
	\label{fig:layered-graph}
\end{figure}

\paragraph{\boldmath Adding edges to $H$} We add two groups of edges to $H$. The first group
	will be called \emph{clique edges} and are added as follows.
	Let $A_i$ be the set of all copies of vertex $v^*_i$ in layer $L_a$.
	Add edges to $H$ such that the vertex set $A_i$ is a clique.
	Similarly, let $B_i$ be the set of all copies of vertex $v^*_i$ in layer $L_b$, 
	add edges to $H$ so that $B_i$ is a clique.  Repeat for all $v^*_i$
Note that clique edges are intra-layer edges.

The second group of edges will be called \emph{free edges} and are added as follows.
For each edge $e^* = (v^*_x, v^*_{y})$ of the dual graph $G^*$, we add a set of 
edges $E_{xy}$ as follows depending on whether or not $e^*$ \emph{crosses} path $\pi_{st}$.
We have the following two cases.
\begin{enumerate}
	\item \emph{Edge $e^*$ does not cross $\pi_{st}$}:~ For every $j \in \sigma(e^*)$, add 
												 edges $(a_x^j, a_y^j)$ and  $(b_x^j, b_y^j)$.
												 Note that all edges added in this case are also \emph{intra-layer} 
												 edges. (See Figure~\ref{fig:layered-graph}(b))

	\item \emph{Edge $e^*$ crosses $\pi_{st}$}:~ For every $j \in \sigma(e^*)$, add 
												 edges $(a_x^j, b_y^j)$ and  $(a_y^j, b_x^j)$.
												 Note that all edges added in this case are \emph{inter-layer} 
												 edges. (See Figure~\ref{fig:layered-graph}(c))
\end{enumerate}

\paragraph{Assigning weights} All \emph{free edges} are assigned a weight of zero.
	Next, we assign weight to \emph{clique edges}. Recall that \emph{clique edge} are of the form
	$(p^k_i, p^\ell_i)$ where $p \in \{a, b\}$ and  $k, \ell \in \sigma(v_i^*)$.
	First, we make these edges directed by adding two directed edges $(p^k_i \rightarrow p^\ell_i)$
	and $(p^k_i \leftarrow p^\ell_i)$. We assign the weights as
	$\wt(p^k_i \rightarrow p^\ell_i) = \wt_\ell$ and 
	$\wt(p^k_i \leftarrow p^\ell_i) = \wt_k$.
	(See also Figure~\ref{fig:layered-graph}.)
	Intuitively, we can think of this assignment of weights as follows: we only pay for a color when 
	entering its vertex -- all consecutive usage is free.

\subsection{\boldmath Min-Color Separator as Shortest Path on $H$}
	In order to cast min-color separator as a shortest path problem on $H$, we need to
	make one final change: for each vertex $v_i^* \in G^*$ we add a pair of \emph{source-sink} nodes
	and connect them to existing nodes of $H$.
	Specifically, for each $v_i^* \in G^*$, we add two special vertices $u_i^a$  (source) and 
	$u_i^b$ (sink) to $H$. Next, we add the source edges $(u_i^a \rightarrow a^k_i)$ for all $k \in \sigma(v_i^*)$ 
	and assign its weight to be $\wt_k$. Similarly, we add the sink edges $(b^k_i \rightarrow u_i^b)$ 
	and assign their weights to be zero. 

	\begin{lemma}
		\label{lemma:shortest-path-H}
		Let $S$ be the min-color separator and $\pi$ be the shortest of all source-sink paths in $H$.
		Then $\wt(S) = \wt(\pi)$.
	\end{lemma}
	\begin{proof}
		Given a color separator $S$, we incrementally build a path $\pi : u_x^a \leadsto u_x^b$ in $H$ as follows.
		Using Lemma~\ref{lemma:well-behaved}, we first obtain a separating cycle $\gamma$ and its well-behaved labeling $\lambda$.
		Fix any vertex $v_x^*$ on $\gamma$ and let $k = \lambda(v^*_x)$.
		Then we add the \emph{source} edge $u_x^a \rightarrow a_x^k$ to $\pi$.
		Now, let $e^*=(v^*_x, v^*_y)$ be the edge adjacent to $v^*_x$ on cycle $\gamma$ in clockwise order. We have two cases.
		\begin{enumerate}
			\item $\lambda(v_x^*) = \lambda(v^*_y) = k$.~ If $e^*$ crosses the reference path $\pi_{st}$, we add the inter-layer 
				edge $a_x^k \rightarrow b_y^k$ to $\pi$. Otherwise, we add intra-layer edge $a_x^k \rightarrow a_y^k$ to $\pi$.
			\item $\lambda(v^*_x) = k$ and  $\lambda(v^*_y) = \ell$.~ We know from Observation~\ref{obs:switch-at-face} that $\ell \in \sigma(v^*_x)$.
					Therefore, we can add the clique edge $a_x^k \rightarrow a_x^\ell$ followed by the free edge $ a_x^\ell \rightarrow  a_y^\ell$ 
					(or $a_x^\ell \rightarrow  b_y^\ell$) to our path $\pi$.
		\end{enumerate}
		We repeat this for all edges of $\gamma$ in clockwise order and finally add the \emph{sink} edge $b_x^k \rightarrow u_x^b$ to the
		path $\pi$. Since $\gamma$ crosses $\pi_{st}$ and odd number of times, we can verify that $\pi$ will be connected.
		Moreover, since $\gamma$ is well-behaved, we pay for each color exactly once and therefore $\wt(\pi) = \wt(\gamma) \leq \wt(S)$.

		For the other direction, given a source-sink path $\pi : u_i^a \leadsto u_i^b$ in $H$, 
		we obtain a separating cycle $\gamma$ and its labeling $\lambda$ as follows.
		First, we simply replace each vertex of the form $a_x^y$ (or $b_x^y$) by its corresponding vertex $v^*_x \in G^*$.
		This gives us a closed walk $W^*$. We claim that all occurrences of any given vertex $v_x^*$ on $W^*$ are consecutive.
		This holds because $\pi$ is a shortest path, so it does not contain a subpath of the form $a_x^k \leadsto a_y^\ell$ (or $b_y^\ell$) $\leadsto a_x^\ell$. 
		For the same reason, $\pi$ will also not contain a subpath of the form $a_x^k \leadsto b_x^\ell$ for some $x \neq i$ (else the source-sink 
		pair $u_x^a \leadsto u_x^b$ would be the shortest.)
		Therefore, we can compress consecutive occurrences of the same vertex $v_x^*$ on $W^*$ into one to obtain the cycle $\gamma$.
		We choose the label $\lambda(v_x^*)$ to be any color $k$ such that $a_x^k$ (or $b_x^k$) lies on $\pi$.
		This gives $w(\gamma) \leq \wt(\pi)$.
		Finally, since  $u_i^a$ and $u_i^b$ lie in different layers, $\pi$ must have taken an odd number of inter-layer edges,
		and therefore it must cross $\pi_{st}$ an odd number of times.
		Since $\gamma$ is a separating cycle, the set of colors $S$ in $\lambda(\gamma)$ form a color separator
		with $\wt(S) \leq \wt(\pi)$.
	\end{proof}

	From Lemma~\ref{lemma:shortest-path-H}, we obtain the following theorem.
	\begin{theorem}
		A  min-color separator on color-connected planar graphs can be computed in polynomial time.
	\end{theorem}

It turns out that color-connectivity is crucial for a polynomial time algorithm for computing
	a min-color separator. Without color connectivity, the problem can easily be shown to be 
\np-hard even on planar graphs of treewidth two, by  a simple reduction from the {\sc Hitting Set} problem. 
	
\end{section}

\section{Hardness of Approximation}
\label{sec:lb-improved}
In this section, we give improved lower bounds for approximating min-color paths when $G$ is not a color-connected planar 
graph. In particular, we show that under plausible complexity theoretic assumptions, without planarity and color-connectivity, 
the \mcp problem admits no polynomial time $O(m^{1-\epsilon})$-approximation algorithm, and no polynomial time 
$O(n^{1/4 -\epsilon})$-approximation algorithm, where $m$ is the number of colors and $n$ is the number of vertices 
in the graph. In fact, these bounds hold even when $G$ has constant treewidth.

We begin by discussing \dvrconj{}, the complexity-theoretic assumption that our lower bounds are based on. A hypergraph $\cg = (X, H)$ over a set of vertices $X$ is \emph{$r$-uniform} if every hyperedge of $H$ has cardinality $r$. For positive integers $n$ and $r$ and real $0 \leq p \leq 1$, the $\cg(n,p,r)$ {\em model} is the probability distribution on $r$-regular $n$-vertex hypergraphs where every subset of the vertex set of size $r$ is included in the set of hyperedges independently with probability $p$.

For a hypergraph $\cg = (X, H)$ and vertex set $X' \subseteq X$ the sub-hypergraph of $\cg$ {\em induced by} $X'$ is denoted by $\cg[X']$ and defined as $(X', \{E \in H ~:~ E \subseteq X'\})$. Our hardness results are based on the assumed hardness of the densest-$k$-subhypergraph problem. Here the input is a hypergraph $\cg$ and integer $k$, and the task is to find the subset $X'$ of $X$ that maximizes the number of hyperedges in $\cg[X']$. This problem is conjectured to be very hard to approximate. Indeed, it is conjectured that there is no polynomial time algorithm that can distinguish between hypergraphs with a dense subhypergraph on $k$ vertices, and hypergraphs drawn from $\cg(n,p,r)$  where the parameters are set in such a way that with high probability each set on $k$ vertices induces a subhypergraph with very few edges. We now formally define this conjecture. 

\begin{Conjecture}[\dvrconj~\cite{chlamtavc2017minimizing}]\label{conj:dvr}
For all constant $r$ and $0 < \alpha,\beta < r-1$, sufficiently small $\eps > 0$, and function $k : \mathbb{N} \rightarrow  \mathbb{N}$ so that $k(n)$ grows polynomially with $n$, $(k(n))^{1+\beta} \leq n^{(1+\alpha)/2}$, there does not exist an algorithm {\sf ALG} that takes as input an $r$-regular $n$-vertex hypergraph $\cg$, runs in polynomial time, and outputs either {\sf dense} or {\sf sparse}, such that:
\begin{itemize}\setlength\itemsep{-.7mm}
\item For every subhypergraph $\cg$ that contains an induced sub-hypergraph on $k = k(n)$ vertices and  $k^{1+\beta}$ hyperedges, ${\sf ALG}(\cg)$ outputs {\sf dense} with high probability.
\item If $\cg$ is drawn from $\cg(n, p, r)$ with $p = n^{\alpha - (r - 1)}$ then ${\sf ALG}(\cg)$ outputs {\sf sparse} with high probability.
\end{itemize}
\end{Conjecture}

In Conjecture~\ref{conj:dvr} and the remainder of this section, {\em with high probability} (w.h.p) means with probability at least $1-O(n^{-c})$ for some constant $c > 0$. In the {\sf dense} case the probability is taken only over the random bits drawn by the algorithm {\sf ALG} if {\sf ALG} is a randomized algorithm. In the  {\sf sparse} case the probability taken over both the draw of $\cg$ from $\cg(n,p,r)$ and the random bits of the algorithm.

We remark that Conjecture~\ref{conj:dvr} is not stated in exactly this way by~\cite{chlamtavc2017minimizing}, indeed their statement of the conjecture leaves some details to interpretation. The statement of the conjecture here is (in our opinion) the ``weakest reasonable formalization'' of the conjecture of~\cite{chlamtavc2017minimizing}, in the sense that every reasonable way to disambiguate their conjecture leads to a statement that implies ours.

			

In order to obtain hardness guarantees for our problem using Conjecture~\ref{conj:dvr}, we will describe a reduction {\sf Red}  that given a hypergraph ${\cal G}$ produces an instance $(G, \sigma)$ of \mcp. We shall then argue that, the images of dense instances under this transformation will have (with high probability ) optimum {\em at most} $x^*_d$, while the images of random instances under this transformation will have optimum at least $x^*_r$, where $x^*_r$ is much bigger than $x^*_d$. We shall call $\rho = x^*_r/ x^*_d $ the \emph{distinguishing ratio} of the reduction. An approximation algorithm for \mcp with ratio smaller than $\rho$ can now (with high probability) distinguish between the images of dense and random instances, thereby refuting Conjecture~\ref{conj:dvr}. This immediately yields the following lemma.

\begin{lemma}\label{lemma:distinguishing-ratio}
Suppose there exists a construction with distinguishing ratio $\rho$, then, assuming Conjecture~\ref{conj:dvr}, there is no polynomial time approximation algorithm for \mcp with approximation ratio less than $\rho$. 
\end{lemma}

In light of Lemma~\ref{lemma:distinguishing-ratio} we will now provide two constructions, one with distinguishing ratio $m^{1-\epsilon}$ where $m$ is the number of colors in the instance $(G, \sigma)$, and the other with distinguishing ratio $|V(G)|^{1-\epsilon}$, implying the previously claimed hardness of approximation results. The constructions for the two cases are identical, however the choices of parameters $\alpha$, $\beta$, $k$ and $r$ are different for the two results. For this reason we will give a more general construction (for a range of the parameter values) and prove some of its properties, and then obtain our results by instantiating the parameters.

We now prove the following bound on the number of vertices in a subhypergraph of $\cg(n, p, r)$
which will be used later.

\begin{lemma}\label{lemma:random-subhypergraph}
Let $\cg$ be drawn from $\cg(n, p, r)$. Then, with high probability, every subhypergraph of $\cg$ with $q = n^{\Omega(1)}$ hyperedges contains $\tilde{\Omega}(\min\{q, (q/p)^{1/r}\})$ vertices. Here $\tilde{\Omega}$ ignores logarithmic factors.
\end{lemma}
\begin{proof}
Define $z = \min\left\{\frac{q \ln \ln n}{3\ln n}, ~\left(\frac{q}{ep\ln n}\right)^{1/r}\right\}$, where 	$\ln$ denotes natural logarithm. We will show that for every fixed set of $z$ vertices, the probability that the sub-hypergraph induced by these vertices contains at least $q$ edges is small. 
Let $H$ be any subhypergraph of $\cg$ with $z$ vertices. 	The probability that $H$ has $q$ edges is the same as probability of $q$ successes in $N = {z \choose r}$ Bernoulli trials where each trial has success probability $p$. Therefore, we have:
	\begin{align*}
		&\text{Pr~[$H$ contains $q$ edges]} \\
            &~~~=~ \binom{N}{q} \cdot p^q \cdot (1-p)^{(N - q)} \\
			&~~~\leq~ \left(\frac{eN}{q}\right)^q \cdot p^q &&\text{since $(1-p)^{(N - q)} \leq 1$}\\
			&~~~\leq~ \left( \frac{ez^r}{q}\right)^q \cdot p^q &&\text{since $N < z^r$}\\
			&~~~\leq~ \left(\frac{ep}{q} \cdot  z^r\right)^q \\
			&~~~\leq~ \left(\frac{1}{\ln n}\right)^q &&\text{since $z \leq \left(\frac{q}{ep \ln n}\right)^{1/r}$}\\
			&~~~\leq~ \left(e^{-\ln \ln n}\right)^q \\
			&~~~\leq~ \left(e^{-\ln \ln n}\right)^{3z \ln n/\ln \ln n} &&\text{since $z \leq \frac{q \ln \ln n}{3\ln n}$} \\
			&~~~\leq~ e^{\ln n^{-3z}} ~~\leq~~ \frac{1}{n^{3z}}
	\end{align*}
Applying the union bound over all possible subsets on $z$ vertices we get that the probability that there exists a sub-hypergraph of $\cg$ on $z$ vertices with at least $q$ hyperedges is at most $n^z \cdot n^{-3z} = 1/n^{2z}$. 
\end{proof}

\paragraph{Construction.}
For every fixed $r$, $\alpha$ and $\beta$ and $k : \mathbb{N} \rightarrow \mathbb{N}$ satisfying the conditions of Conjecture~\ref{conj:dvr} we describe a reduction that, given a $r$-uniform hypergraph $\cg = (X, H)$ on $n$ vertices constructs an instance of \mcp, namely a graph $G=(V, E)$ and a coloring function $\sigma : V \rightarrow 2^X$. Note that the vertex-set $X$ of the hypergraph forms the color set for $(G, \sigma)$. We set the following parameters: $q = k^{1+\beta}$, and $\ell = \frac{q}{(r+1)\ln n}$. Observe that because $k$ grows polynomially with $n$, so does $q$. Thus $q$ satisfies the conditions of Lemma~\ref{lemma:random-subhypergraph}. The construction proceeds as follows.

\begin{figure}[h!]
	\centering
	\includegraphics{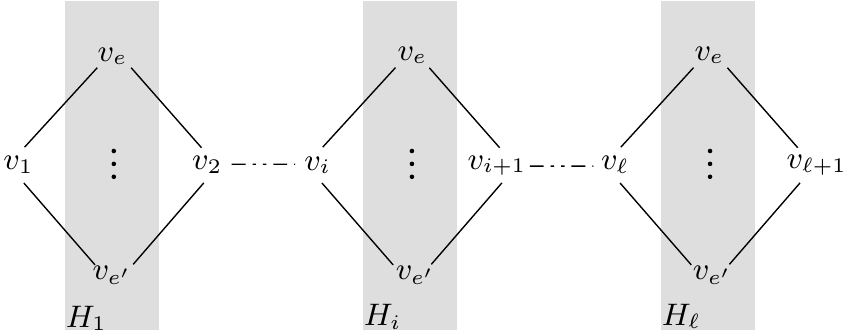}
	\caption{An example of the construction. The groups $H_1, H_i$ and $H_\ell$ are shown shaded in gray.}
	\label{fig:lb-vertex-colored}
\end{figure}

\begin{itemize}
	\item Add $\ell+1$ vertices $v_1, v_2, \dots, v_{\ell+1}$ to $G$ and arrange them sequentially in the plane
				from left to right. (See also Figure~\ref{fig:lb-vertex-colored}.)
	\item Uniformly partition hyperedges $H$ into $\ell$ groups as $H_1, H_2, \dots, H_\ell$. That is,
				every hyperedge is assigned a group with a probability $1/\ell$ independent of other hyperedges.
	\item For each hyperedge $e \in H_i$, add a vertex $v_e$ and connect it to vertices $v_{i}$ and $v_{i+1}$.
				Assign $\sigma(v_e) = \adj(j)$, where $\adj(e)$ to denote vertices adjacent to hyperedge $e \in H$.
	\item Let $s = v_1$ be the source $t = v_{\ell+1}$ be the destination. 
\end{itemize}

The following lemma follows immediately from a standard ``balls and bins'' kind of argument, using that we set $\ell = \frac{q}{(r+1)\ln n}$. 

\begin{lemma}\label{lemma:nonempty-groups}
With high probability, for every subset $H^* \subseteq H$ of $q$ hyperedges of $\cg$, every group $H_i \in \{H_1, H_2, \dots, H_\ell\}$ contains at least one edge from $H^*$. Here $n$ is the number of vertices of $\cg$.
\end{lemma}
\begin{proof}
Fix a subset $H^*$ on $q$ hyperedges. We say that group $H_i$ is \emph{$H^*$-empty} if $H_i \cap H^* = \emptyset$ (in other words $H_i$ contains no edge from $H^*$). The probability $H_i$ is $H^*$-empty is upper by $(1 - \frac{1}{\ell})^q \leq \frac{1}{n^{r+1}}$. A union bound over all $\ell \leq n^r$ groups $H_i$ proves the statement of the lemma. 
\end{proof}


\begin{lemma}\label{lemma:dense-random-values}
Let $(G, \sigma)$ be the result of applying the transformation above to a hypergraph $\cg$. 
\begin{enumerate}
\item If $\cg$ contains a sub-hypergraph on $k$ vertices and $q$ hyperedges, then with high probability $G$ contains an $s-t$ path that uses at most $k$ colors. 
\item If $\cg$ is drawn from $\cg(n, p, r)$ with $p = n^{\alpha - (r - 1)}$ then with high probability every $s-t$ path in $G$ uses at least $\tilde{\Omega}(\min\{q, (q/p)^{1/r}\})$ colors.
\end{enumerate}
\end{lemma}

\begin{proof}
For the first point, $\cg$ contains a subhypergraph with $k$ vertices and at least $q = k^{\beta+1}$ hyperedges. Let $H^*$ be the set of these hyperedges. Applying Lemma~\ref{lemma:nonempty-groups}, we get that each of the $\ell$ groups $H_i$ contain at least one edge from $H^*$. Therefore, the colored graph $G$ contains a path that only uses colors corresponding to vertices in hyperedges in the edge set $H^*$. Since the number of such vertices is at most $k$, there exists a path in $G$ with at most $k$ colors.

For the second point, the set of colors of any $s-t$ path $P$ in $G$ corresponds to a subhypergraph of $\cg$ with $\ell$ edges. Applying Lemma~\ref{lemma:random-subhypergraph} on this subhypergraph and observing that $\ell = \tilde{\Omega}(q)$, yields that $P$ uses at least $\tilde{\Omega}(\min\{q, (q/p)^{1/r}\})$ colors.
\end{proof}

From Lemma~\ref{lemma:dense-random-values}, we obtain the following lower bound on the distinguishing ratio:
\begin{align}
\rho ~~&\geq~~ \frac{\min\{k^{\beta+1},~~ (k^{\beta+1}/n^{1+\alpha - r})^{1/r}\}}{k} \nonumber \\
    &=~~ \min\left\{k^\beta,~~ \left(\frac{k^{\beta+1-r}}{n^{1+\alpha - r}}\right)^{1/r}\right\} \label{eqn:distinguishing-ratio}
\end{align}

We are now ready to prove the first of our hardness of approximation bounds. 

\begin{lemma}\label{lem:hardcolors} 
Assuming \dvrconj, \mcp on a graph $G = (V, E)$ with a coloring function $\sigma : V \rightarrow 2^{[m]}$ cannot be 
approximated within a factor of $m^{1-\epsilon}$, for any $\epsilon > 0$.
\end{lemma}

\begin{proof}
We choose the parameters $\alpha = \sqrt{r}-1$, $\beta = \alpha - \epsilon'$ and $k = n^{\frac{1}{\sqrt{r}+1}}$ for some sufficiently small $\epsilon' > 0$. It is easy to verify that these parameter values satisfy the requirements for Conjecture~\ref{conj:dvr}.That is, we have $0 \leq \alpha, \beta < r-1$ and $k^{\beta+1} = n^{\frac{\sqrt{r} - \eps}{\sqrt{r}+1}} \leq n^{1+\alpha}$.

Substituting the values in Equation~\ref{eqn:distinguishing-ratio}, we obtain the following bound for the distinguishing ratio. 
\begin{align*}
k^\beta ~~=~~ n^{\frac{\sqrt{r}-1 -\eps}{\sqrt{r}+1}} ~~&=~~  n^{1 - \frac{2 +\eps}{\sqrt{r}+1}} ~~=~~ n^{1-\eps'} \\ 
											\text{~~~where $\eps' = \frac{2 +\eps}{\sqrt{r}+1}$} \\
		\text{Similarly, we have~~} \\
\left(\frac{k^{\beta+1-r}}{n^{1+\alpha - r}}\right)^{1/r} ~~&=~~ n^{\frac{\sqrt{r} - 1 - \eps/r}{\sqrt{r} + 1}} ~~=~~ n^{1-\eps''} \\
										    \text{~~~where $\eps'' = \frac{2 +\eps/r}{\sqrt{r}+1}$}
\end{align*}
Setting $\eps'$ sufficiently small and $r$ sufficiently large (but independent of $n$) yields that the distinguishing ratio $\rho$ is at least $n^{1-\eps}$. Here $n$ is the number of vertices in $\cg$, which is equals to the number $m$ of colors in $(G, \sigma)$. The statement now follows from the lower bound on the distinguishing ratio together with Lemma~\ref{lemma:distinguishing-ratio}.
\end{proof}



The above lemma shows that, it is quite unlikely to find a good approximation in terms of the number of colors. We will now obtain a bound in terms of number of vertices of $G$. Observe that the number of vertices in $G$ is $h+\ell = \Theta(h)$ where
$h$ is the number of hyperedges of $\cg$. We can rewrite Equation~\ref{eqn:distinguishing-ratio} by expressing the probability $p = n^{1+\alpha-r}$ in terms of the expected number $h = \theta(n^{1+\alpha})$ of edges in a hypergraph drawn from $\cg(n, p, r)$. Note that the actual number of edges of a hypergraph $\cg$ drawn from $\cg(n, p, r)$ is $\theta(n^{1+\alpha})$ with high probability. Thus, we can rewrite the lower bound on the distinguishing ratio from Equation~\ref{eqn:distinguishing-ratio} in terms of $h$ as follows, using that $p = h^{(1+\alpha-r) / (1+\alpha)}$.

\begin{align}
	\rho ~~\geq~~ \min\left\{k^\beta,~~ \left(\frac{k^{\beta+1-r}}{h^{(1+\alpha - r)/(1+\alpha)}}\right)^{1/r}\right\} \label{eqn:distinguishing-ratio-m}
\end{align}

Moreover, since the number of vertices in the colored graph $G$ is $\Theta(h)$, we obtain the following theorem.
\begin{lemma}\label{lem:hardVertices} 
Assuming \dvrconj, \mcp on a graph $G=(V, E)$ with a coloring function $\sigma : V \rightarrow 2^{[m]}$ 
cannot be $O(|V|^{1/4 -\epsilon})$-approximated, for any $\epsilon > 0$.
\end{lemma}

\begin{proof}
We set the parameters $\alpha =  \frac{r-1}{r+1}$, $\beta = \alpha - \eps$ and $k = h^{\frac{r+1}{4r}}$, and choose $\epsilon' > 0$ sufficiently small. It is easy to verify that these parameter values satisfy Conjecture~\ref{conj:dvr}. More precisely, we have $0 < \alpha, \beta < (r-1)$ and $k^{\beta + 1} ~<~ k^{\alpha + 1} ~=~ h^{\frac{r+1}{4r} \cdot \frac{2r}{r+1}} ~=~ \sqrt{h} ~=~ n^{\frac{1+\alpha}{2}}$. Substituting these values into Equation~\ref{eqn:distinguishing-ratio-m}, we obtain:
\begin{align*}
	k^\beta ~~=~~ h^{\frac{r+1}{4r} \cdot (\frac{r-1}{r+1} - \eps)} ~~&=~~ h^{\frac{1}{4} - \frac{1+\eps(r+1)}{4r}} \\
        ~~=~~ h^{1/4 - \eps'}~~&~~\text{~~~where $\eps' = \frac{1+\eps(r+1)}{4r}$} \\
		\text{Similarly, we have~~}  \\
		\left(\frac{k^{\beta+1-r}}{h^{(1+\alpha - r)/(1+\alpha)}}\right)^{1/r}  ~~&=~~ h^{\frac{1}{4} - \frac{1+\eps(r+1)/r}{4r}}\\
             ~~=~~ h^{1/4-\eps''}~~&~~\text{~~~where $\eps'' = \frac{1+\eps(r+1)/r}{4r}$}
\end{align*}

Setting $\eps'$ sufficiently small and $r$ sufficiently large (but independent of $n$) yields that the distinguishing ratio $\rho$ is at least $h^{\frac{1}{4}-\eps}$. Finally, since the number of vertices in $G$ is $\Theta(h)$ (with high probability), Lemma~\ref{lemma:distinguishing-ratio} completes the proof. 
\end{proof}

Observe that the \mcp instances $G$ constructed in Lemmata~\ref{lem:hardVertices} and~\ref{lem:hardcolors} are planar and have treewidth two. One can make color-connected (but non-planar) equivalent instances by adding a single vertex $v^*$ with color set $\sigma(v^*) = [m]$ and connecting it to all other vertices. This only increases treewidth of $G$ by $1$. We summarize our results in the following theorem. 

\hardnessmain*
\section{Conclusion}
\label{sec:conclusion}
In this paper we gave the first polynomial-time constant factor approximation algorithm for the \mcp problem on 
color-connected planar graphs, answering an open question posed in~\cite{bandyapadhyay2018improved,ChanK14}. This algorithm immediately yields a constant factor approximation algorithm for the {\sc Barrier Resilience} and {\sc Minimum Constraint Removal} problems.  In fact we obtained a constant factor approximation  for a substantially more general {\sc Minimum Color Prize Collecting Steiner Forest} version of the problem, which  generalizes classic {\sc Steiner Forest} and {\sc Prize-Collecting Steiner Forest} problems on planar graphs. We complemented our algorithmic findings by showing that neither the assumption on planarity nor the connectivity of colors can be dropped from our results; either of these would lead to strong inapproximability results.

 We believe that Theorem~\ref{thm:approxminmax} together with the methods for rounding algorithm for \lp will find further applications for obstacle removal problems. It is interesting both to pin down the “right” constant factor approximation ratio for \mcp for planar color connected graphs, as well as evaluate the performance of heuristic algorithms for \mcp based on rounding \lp on practical instances.




\bibliographystyle{siam}
\bibliography{refs}
\end{document}